\documentclass[twocolumn]{aastex63}
\usepackage{natbib, color,amsmath}
\usepackage[normalem]{ulem}
\usepackage{epstopdf}
\usepackage{graphicx}
\definecolor{red}{rgb}{1.0,0.0,0.0}

\newcommand{\Mj}[1]{$M_\mathrm{Jup}$}

\usepackage[utf8]{inputenc}

\newcommand{\target}{$\beta$ Pic}

\begin{document}
\title{The Gemini Planet Imager Exoplanet Survey: Dynamical Mass of the Exoplanet $\beta$~Pictoris~b from Combined Direct Imaging and Astrometry}

\correspondingauthor{Eric L. Nielsen}
\email{enielsen@stanford.edu}

\author[0000-0001-6975-9056]{Eric L. Nielsen}
\affiliation{Kavli Institute for Particle Astrophysics and Cosmology, Stanford University, Stanford, CA 94305, USA}

\author[0000-0002-4918-0247]{Robert J. De Rosa}
\affiliation{Kavli Institute for Particle Astrophysics and Cosmology, Stanford University, Stanford, CA 94305, USA}

\author[0000-0003-0774-6502]{Jason J. Wang}
\altaffiliation{51 Pegasi b Fellow}
\affiliation{Department of Astronomy, California Institute of Technology, Pasadena, CA 91125, USA}

\author[0000-0001-9525-3673]{Johannes Sahlmann}
\affiliation{Space Telescope Science Institute, Baltimore, MD 21218, USA}

\author[0000-0002-6221-5360]{Paul Kalas}
\affiliation{Department of Astronomy, University of California, Berkeley, CA 94720, USA}
\affiliation{SETI Institute, Carl Sagan Center, 189 Bernardo Ave.,  Mountain View CA 94043, USA}
\affiliation{Institute of Astrophysics, FORTH, GR-71110 Heraklion, Greece}

\author[0000-0002-5092-6464]{Gaspard Duch\^ene}
\affiliation{Department of Astronomy, University of California, Berkeley, CA 94720, USA}
\affiliation{Univ. Grenoble Alpes/CNRS, IPAG, F-38000 Grenoble, France}

\author[0000-0003-0029-0258]{Julien Rameau}
\affiliation{Univ. Grenoble Alpes/CNRS, IPAG, F-38000 Grenoble, France}
\affiliation{Institut de Recherche sur les Exoplan{\`e}tes, D{\'e}partement de Physique, Universit{\'e} de Montr{\'e}al, Montr{\'e}al QC, H3C 3J7, Canada}

\author[0000-0002-5251-2943]{Mark S. Marley}
\affiliation{NASA Ames Research Center, MS 245-3, Mountain View, CA 94035, USA}

\author{Didier Saumon}
\affiliation{Los Alamos National Laboratory, P.O. Box 1663, Los Alamos, NM 87545, USA}

\author[0000-0003-1212-7538]{Bruce Macintosh}
\affiliation{Kavli Institute for Particle Astrophysics and Cosmology, Stanford University, Stanford, CA 94305, USA}

\author[0000-0001-6205-9233]{Maxwell A. Millar-Blanchaer}
\altaffiliation{NASA Hubble Fellow}
\affiliation{Jet Propulsion Laboratory, California Institute of Technology, Pasadena, CA 91109, USA}

\author[0000-0002-9350-4763]{Meiji M. Nguyen}
\affiliation{Department of Astronomy, University of California, Berkeley, CA 94720, USA}

\author[0000-0001-5172-7902]{S. Mark Ammons}
\affiliation{Lawrence Livermore National Laboratory, Livermore, CA 94551, USA}

\author[0000-0002-5407-2806]{Vanessa P. Bailey}
\affiliation{Jet Propulsion Laboratory, California Institute of Technology, Pasadena, CA 91109, USA}

\author[0000-0002-7129-3002]{Travis Barman}
\affiliation{Lunar and Planetary Laboratory, University of Arizona, Tucson AZ 85721, USA}

\author{Joanna Bulger}
\affiliation{Institute for Astronomy, University of Hawaii, 2680 Woodlawn Drive, Honolulu, HI 96822, USA}
\affiliation{Subaru Telescope, NAOJ, 650 North A{'o}hoku Place, Hilo, HI 96720, USA}

\author[0000-0001-6305-7272]{Jeffrey Chilcote}
\affiliation{Department of Physics, University of Notre Dame, 225 Nieuwland Science Hall, Notre Dame, IN, 46556, USA}

\author[0000-0003-0156-3019]{Tara Cotten}
\affiliation{Department of Physics and Astronomy, University of Georgia, Athens, GA 30602, USA}

\author{Rene Doyon}
\affiliation{Institut de Recherche sur les Exoplan{\`e}tes, D{\'e}partement de Physique, Universit{\'e} de Montr{\'e}al, Montr{\'e}al QC, H3C 3J7, Canada}

\author[0000-0002-0792-3719]{Thomas M. Esposito}
\affiliation{Department of Astronomy, University of California, Berkeley, CA 94720, USA}

\author[0000-0002-0176-8973]{Michael P. Fitzgerald}
\affiliation{Department of Physics \& Astronomy, University of California, Los Angeles, CA 90095, USA}

\author[0000-0002-7821-0695]{Katherine B. Follette}
\affiliation{Physics and Astronomy Department, Amherst College, 21 Merrill Science Drive, Amherst, MA 01002, USA}

\author[0000-0003-3978-9195]{Benjamin L. Gerard}
\affiliation{University of Victoria, 3800 Finnerty Rd, Victoria, BC, V8P 5C2, Canada}
\affiliation{National Research Council of Canada Herzberg, 5071 West Saanich Rd, Victoria, BC, V9E 2E7, Canada}

\author[0000-0002-4144-5116]{Stephen J. Goodsell}
\affiliation{Gemini Observatory, 670 N. A'ohoku Place, Hilo, HI 96720, USA}

\author{James R. Graham}
\affiliation{Department of Astronomy, University of California, Berkeley, CA 94720, USA}

\author[0000-0002-7162-8036]{Alexandra Z. Greenbaum}
\affiliation{Department of Astronomy, University of Michigan, Ann Arbor, MI 48109, USA}

\author[0000-0003-3726-5494]{Pascale Hibon}
\affiliation{Gemini Observatory, Casilla 603, La Serena, Chile}

\author[0000-0003-1498-6088]{Li-Wei Hung}
\affiliation{Natural Sounds and Night Skies Division, National Park Service, Fort Collins, CO 80525, USA}

\author{Patrick Ingraham}
\affiliation{Large Synoptic Survey Telescope, 950N Cherry Ave., Tucson, AZ 85719, USA}

\author[0000-0002-9936-6285]{Quinn Konopacky}
\affiliation{Center for Astrophysics and Space Science, University of California San Diego, La Jolla, CA 92093, USA}

\author[0000-0001-7687-3965]{James E. Larkin}
\affiliation{Department of Physics \& Astronomy, University of California, Los Angeles, CA 90095, USA}

\author{J\'er\^ome Maire}
\affiliation{Center for Astrophysics and Space Science, University of California San Diego, La Jolla, CA 92093, USA}

\author[0000-0001-7016-7277]{Franck Marchis}
\affiliation{SETI Institute, Carl Sagan Center, 189 Bernardo Ave.,  Mountain View CA 94043, USA}

\author[0000-0002-4164-4182]{Christian Marois}
\affiliation{National Research Council of Canada Herzberg, 5071 West Saanich Rd, Victoria, BC, V9E 2E7, Canada}
\affiliation{University of Victoria, 3800 Finnerty Rd, Victoria, BC, V8P 5C2, Canada}

\author[0000-0003-3050-8203]{Stanimir Metchev}
\affiliation{Department of Physics and Astronomy, Centre for Planetary Science and Exploration, The University of Western Ontario, London, ON N6A 3K7, Canada}
\affiliation{Department of Physics and Astronomy, Stony Brook University, Stony Brook, NY 11794-3800, USA}

\author[0000-0001-7130-7681]{Rebecca Oppenheimer}
\affiliation{Department of Astrophysics, American Museum of Natural History, New York, NY 10024, USA}

\author{David Palmer}
\affiliation{Lawrence Livermore National Laboratory, Livermore, CA 94551, USA}

\author{Jennifer Patience}
\affiliation{School of Earth and Space Exploration, Arizona State University, PO Box 871404, Tempe, AZ 85287, USA}

\author[0000-0002-3191-8151]{Marshall Perrin}
\affiliation{Space Telescope Science Institute, Baltimore, MD 21218, USA}

\author{Lisa Poyneer}
\affiliation{Lawrence Livermore National Laboratory, Livermore, CA 94551, USA}

\author{Laurent Pueyo}
\affiliation{Space Telescope Science Institute, Baltimore, MD 21218, USA}

\author[0000-0002-9246-5467]{Abhijith Rajan}
\affiliation{Space Telescope Science Institute, Baltimore, MD 21218, USA}

\author[0000-0002-9667-2244]{Fredrik T. Rantakyr\"o}
\affiliation{Gemini Observatory, Casilla 603, La Serena, Chile}

\author[0000-0003-2233-4821]{Jean-Baptiste Ruffio}
\affiliation{Kavli Institute for Particle Astrophysics and Cosmology, Stanford University, Stanford, CA 94305, USA}

\author[0000-0002-8711-7206]{Dmitry Savransky}
\affiliation{Sibley School of Mechanical and Aerospace Engineering, Cornell University, Ithaca, NY 14853, USA}

\author{Adam C. Schneider}
\affiliation{School of Earth and Space Exploration, Arizona State University, PO Box 871404, Tempe, AZ 85287, USA}

\author[0000-0003-1251-4124]{Anand Sivaramakrishnan}
\affiliation{Space Telescope Science Institute, Baltimore, MD 21218, USA}

\author[0000-0002-5815-7372]{Inseok Song}
\affiliation{Department of Physics and Astronomy, University of Georgia, Athens, GA 30602, USA}

\author[0000-0003-2753-2819]{Remi Soummer}
\affiliation{Space Telescope Science Institute, Baltimore, MD 21218, USA}

\author{Sandrine Thomas}
\affiliation{Large Synoptic Survey Telescope, 950N Cherry Ave., Tucson, AZ 85719, USA}

\author[0000-0001-5299-6899]{J. Kent Wallace}
\affiliation{Jet Propulsion Laboratory, California Institute of Technology, Pasadena, CA 91109, USA}

\author[0000-0002-4479-8291]{Kimberly Ward-Duong}
\affiliation{Physics and Astronomy Department, Amherst College, 21 Merrill Science Drive, Amherst, MA 01002, USA}

\author{Sloane Wiktorowicz}
\affiliation{Department of Astronomy, UC Santa Cruz, 1156 High St., Santa Cruz, CA 95064, USA }

\author[0000-0002-9977-8255]{Schuyler Wolff}
\affiliation{Leiden Observatory, Leiden University, 2300 RA Leiden, The Netherlands}


\begin{abstract}
We present new observations of the planet $\beta$~Pictoris~b from 2018 with GPI, the first GPI observations following conjunction.  Based on these new measurements, we perform a joint orbit fit to the available relative astrometry from ground-based imaging, the \textit{Hipparcos} Intermediate Astrometric Data (IAD), and the \textit{Gaia} DR2 position, and demonstrate how to incorporate the IAD into direct imaging orbit fits.  We find a mass consistent with predictions of hot-start evolutionary models and previous works following similar methods, though with larger uncertainties: 12.8$^{+5.3}_{-3.2}$~M$_\textrm{Jup}$.  Our eccentricity determination of $0.12^{+0.04}_{-0.03}$ disfavors circular orbits.  We consider orbit fits to several different imaging datasets, and find generally similar posteriors on the mass for each combination of imaging data.  Our analysis underscores the importance of performing joint fits to the absolute and relative astrometry simultaneously, given the strong covariance between orbital elements.  Time of conjunction is well constrained within 2.8 days of 2017 September 13, with the star behind the planet's Hill sphere between 2017 April 11 and 2018 February 16 ($\pm$ 18 days).  Following the recent radial velocity detection of a second planet in the system, $\beta$~Pic~c, we perform additional two-planet fits combining relative astrometry, absolute astrometry, and stellar radial velocities.  These joint fits find a significantly smaller mass for the imaged planet $\beta$~Pic~b, of $8.0\pm2.6$~M$_\textrm{Jup}$, in a somewhat more circular orbit.  We expect future ground-based observations to further constrain the visual orbit and mass of the planet in advance of the release of \textit{Gaia} DR4.
\end{abstract}

\section{Introduction}

Masses of exoplanets detected by the radial velocity method can be directly measured to within $\sin(i)$, as can the mass ratio between microlensing planets and their parent star, and masses can be inferred for transiting planet systems by modeling transit timing variations.  The masses of directly imaged planets, however, must be inferred from evolutionary models if only imaging data are available.  These models predict the mass of the planet as a function of age of the system and luminosity of the planet.  While the COND models \citep{baraffe03} have been consistent with upper limits on directly imaged planet masses \citep{Lagrange:2012,Wang:2018}, direct measurements of the mass allow for a more robust testing of the models.  Giant planets are most easily imaged around young stars ($\lesssim$100 Myr), which tend to be too active for precise radial velocity measurements (e.g., \citealt{Lagrange:2012} describe searching for a $\sim$10 m/s signal in RV data with a $\sim$3 km/s peak-to-peak RV variation).  Astrometry, however, is less affected by stellar activity, and represents a way forward to determining the dynamical mass of these planets from stellar reflex motion.

In particular, the second \textit{Gaia} data release (DR2)  gives independent measurements of $\sim$2016 position and proper motions for $\sim$1 billion stars.  Recently, \citet{snellen:2018} combined an orbit fit to direct imaging data by \citet{Wang:2016gl} with \textit{Hipparcos} Intermediate Astrometric Data and \textit{Gaia} positions for the planet $\beta$ Pictoris b.  This combination of the orbital period from imaging, with absolute positions in $\sim$1991 and $\sim$2016 resulted in a measurement of the planet mass of 11$\pm$2 M$_{\rm Jup}$.  A similar analysis was undertaken by \citet{dupuy:2019} earlier this year.

$\beta$ Pic is a young, nearby ($d$ = 19.44 pc), intermediate-mass ($\sim$1.8 M$_\odot$) star that hosts a bright edge-on debris disk \citep{smith:1984,kalas:1995,wahhaj:2003, weinberger:2003,golimowski:2006,Nielsen:2014}.  It is part of the $\beta$ Pic moving group \citep{barrado1999,zucerkman:2001,binks:2014,bell:2015}, which sets the age of the star to 26 $\pm$ 3 Myr \citep{Nielsen:2016ct}.  $\beta$ Pic b was one of the first directly imaged exoplanets, first observed on the north-east side of the star in 2003 \citep{Lagrange:2009hq}, before being confirmed after it passed behind the star to the south-west side \citep{Lagrange:2010}.  Subsequent observations allowed the orbit of the planet to be determined to increasing accuracy \citep{currie:2011,Chauvin:2012,Nielsen:2014,Macintosh:2014js,MillarBlanchaer:2015ha,Wang:2016gl}.  The planet's orbital plane has been found to be very similar to the plane of the disk, and though a transit-like event was observed in 1981 \citep{LecavelierDesEtangs:2009jt}, additional relative astrometry has ruled out the possibility of the planet itself transiting, though the planet's Hill sphere passes in front of the star (e.g., \citealt{MillarBlanchaer:2015ha,Wang:2016gl}).

The Gemini Planet Imager (GPI, \citealt{Macintosh:2014js}) is an extreme adaptive optics system on the Gemini South 8-m telescope optimized for detecting self-luminous giant exoplanets.  $\beta$ Pic was observed multiple times by GPI since 2013, tracking the orbit of the planet as it moved closer to the star \citep{Wang:2016gl}.  Here we present new observations from GPI in 2018, following conjunction, and a joint fit of the imaging data and \textit{Hipparcos} and \textit{Gaia} astrometry, along with an estimate of the mass of the planet.

\section{Observations}

\subsection{New GPI data}\label{sec:gpi_data}
$\beta$ Pic was observed in 2018 after a hiatus in which the planet passed too close to the star in angular projection ($\lesssim$ 0.15"). Due to the close angular separation of the planet and the star, we chose to observe $\beta$ Pic b in $J$-band in order to maximize sensitivity at $\sim150$~mas radius while maintaining a favorable flux ratio of the planet. In this paper, we present two epochs of GPI $J$-band integral field spectroscopy observations of the planet. The first epoch was taken on 2018 September 21 between 8:42 and 10:02 UT. After discarding frames in which the AO loops opened, we obtained a total of 59 exposures with integration times of 60~s. A total of $36.8^{\circ}$ of field rotation was obtained for angular differential imaging \citep[ADI;][]{marois:2006}. The second epoch was taken on 2018 November 18 between 5:51 and 9:13 UT with a total of 145 exposures, each of which is comprised of four co-added 14.5~s frames. These observations were better timed and a total of $96.9^{\circ}$ of field rotation was obtained for angular differential imaging. 

\begin{figure}[ht]
\includegraphics[width=0.5\textwidth]{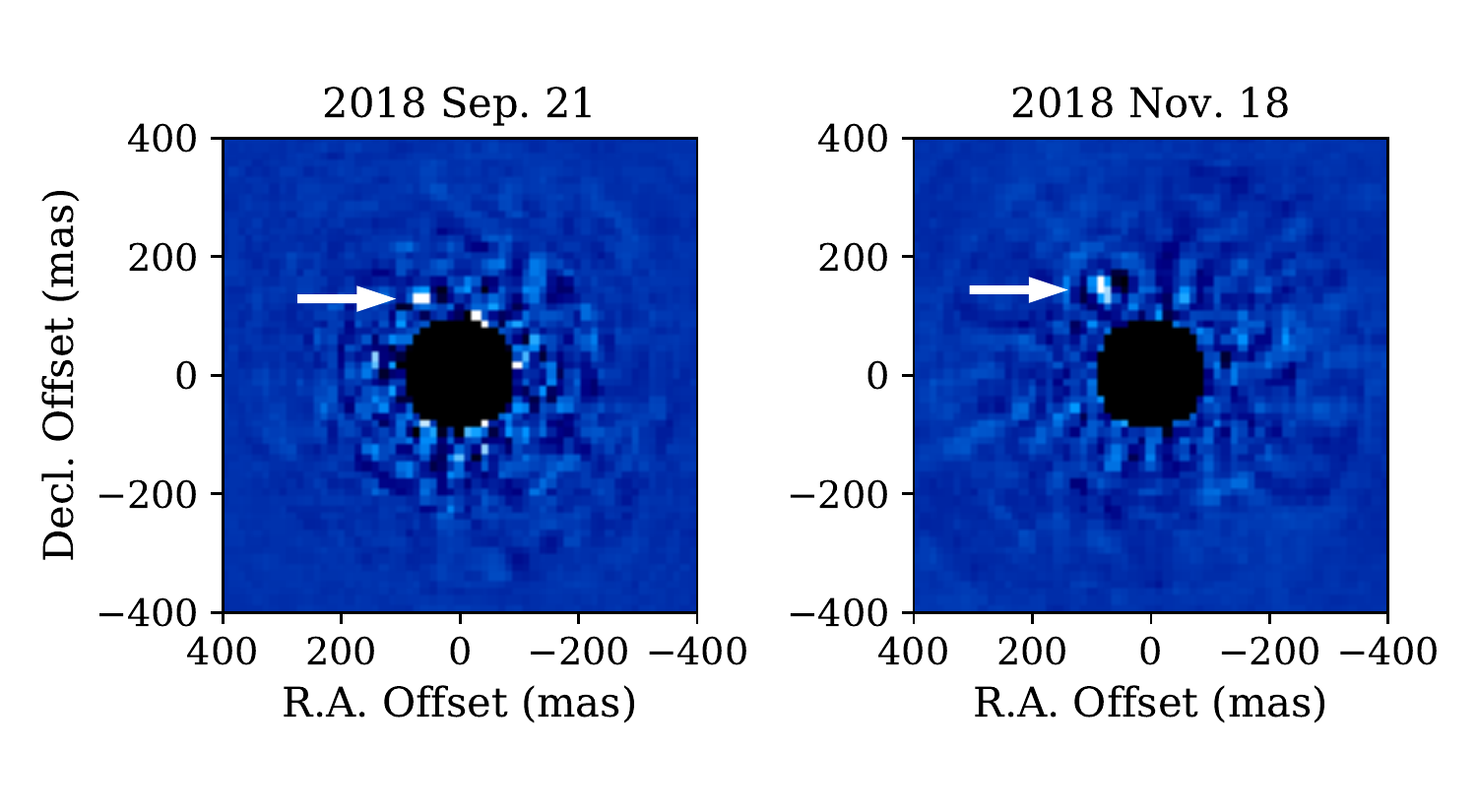}
\caption{GPI images of $\beta$ Pic b processed with the automatic GPIES pipeline. The images are rotated North-up-East-left and have not been flux calibrated. The colors are presented on a linear scale. The white arrow points to the location of the planet.  \label{fig:gpi_data}}
\end{figure}

The data were first reduced using the automated GPIES data reduction pipeline \citep{Wang:2018}, with one notable exception. During the night of September 21, 2018, GPI was not able to access the Gemini Facility Calibration Unit (GCAL) and could not obtain a Argon arc lamp snapshot before each observation sequence to correct for instrument flexure \citep{Wolff:2014}. For the $\beta$ Pic observations, we corrected instrument flexure manually through visual inspection. This did not significantly impact the spatial image reconstruction of the 3-pixel box extraction algorithm used in the GPI Data Reduction Pipeline \citep[DRP;][]{Perrin:2014, Perrin:2016gm}, but it likely affected our spectral accuracy. However, for the purpose of astrometry, we collapse the spectral datacubes into a broadband image, so the impact on astrometry is minimal. In both epochs, we used the satellite spots, four fiducial diffraction spots centered on the location of the star \citep{Sivaramakrishnan2006, Marois2006b}, to locate the star behind the coronagraph in each wavelength slice of each spectral datacube \citep{Wang:2014}. The stellar point spread function (PSF) was then subtracted out using \texttt{pyKLIP} \citep{Wang:2015th}, which uses principal component analysis \citep{Soummer:2012ig, Pueyo:2015} constructed from images taken at other times (ADI) and wavelengths \citep[spectral differential imaging;][]{Sparks2002}. The reductions of the two epochs using 20 principal components to model and subtract out the stellar PSF and averaged over time and wavelength are shown in Figure \ref{fig:gpi_data}. We estimated a signal-to-noise ratio of 4.5 and 11.7 for the September and November datasets respectively.

To measure the position of $\beta$ Pic b in each dataset, we follow the same technique that was outlined in \citet{Wang:2016gl} where the signal of the planet is forward modeled through the data reduction process and the forward model is then fit to the data. In these reductions, we found it was optimal to discard frames from the sequences due to varying image quality. For both datasets, we ordered datacubes by the contrast in each single datacube at 250~mas. For the September 21st epoch, we only used the best 40 datacubes, resulting in a total integration time of 40 minutes. For the November 18th epoch, we used the best 120 frames, resulting in a total integration time of 116 minutes. 

We then used the astrometry modules in \texttt{pyKLIP} to run a stellar PSF subtraction that simultaneously forward models the PSF of the planet. For both epochs, we built 15 principal components from the 150 more correlated reference PSFs, where the reference PSFs are drawn from frames at other wavelengths and times where $\beta$ Pic b moved at least 1 pixel in the image due to a combination of ADI and spectral differential imaging (SDI). For the September 21st epoch, we broke up the image between 6.5 and 25.6 pixels from the star into three concentric annuli of 4.0, 6.7, and 8.4 pixels in width respectively. We then broke each annuli into 4 sectors, and ran our stellar PSF subtraction and forward modeling on each sector. For the November 18th epoch, we only used one annulus centered on the star with an inner radius of 6.5 pixels and an outer radius of 19.2 pixels. We did not split up this annulus. The annuli geometry were defined by the focal plane mask and the edge of the field of view. The planet's position in the data was then fit over a 9-pixel wide box centered on the estimated location of the planet. In this box were pixels that fell inside the focal plane mask, and not included in our reduction. We did not consider these pixels in the fit, reducing the number of data points by a few. The fit was done using the Bayesian framework described in \citet{Wang:2016gl} where we used a Gaussian process to model the correlated speckle noise present in the data. 
Due to the close separation of the planet in these two epochs, we did not trust the assumption of Gaussian noise used in our Bayesian framework when estimating uncertainties on the planet's location. To empirically quantify this and any residual biases in the forward model, we injected simulated planets into the datasets with a spectrum from a model fit to $\beta$ Pic b's spectrum at the same separation as $\beta$ Pic b, but at position angles that are at least 3 full-widths at half-maximum apart from the measured position of the planet. We injected one simulated planet at a time, measured its astrometry, and compared it to the true position we injected it at. We found a scatter in the position of 0.3 pixels for the September 21st epoch and a scatter in the position of 0.13 pixels for the position of the November 18th epoch. We found the average measured astrometry of the simulated planets was biased by $< 0.02$ pixels, so we conclude that fitting biases are negligible. We use the scatter in the simulated planet positions as the uncertainty in the position of $\beta$ Pic b.
To obtain relative astrometry of the host star, we assumed a star centering precision of 0.05 pixels \citep{Wang:2014}, a plate scale value of $14.161 \pm 0.021$~mas/pixel, and a residual North angle correction of $0.45^{\circ} \pm 0.11^{\circ}$ \citep{derosa2019_north}. The relative astrometry is reported in Table \ref{tbl:imaging_data}.

\begin{deluxetable*}{ccccc}
\tablewidth{0pt}
\tabletypesize{\scriptsize}
\tablecaption{Relative astrometry of $\beta$ Pic b\label{tbl:imaging_data}
}
\tablehead{ \colhead{Epoch} & \colhead{Sep (")} & \colhead{PA (deg)} & \colhead{Instrument} & \colhead{Reference}}
\startdata
2008-11-11 & 0.210 $\pm$ 0.027 & 211.49 $\pm$ 1.9 & VLT/NACO & \citet{currie:2011} \\
2003-11-10 & 0.413 $\pm$ 0.022& 34 $\pm$ 4 & VLT/NACO & \citet{Chauvin:2012} \\
2009-10-25 & 0.299 $\pm$ 0.014& 211 $\pm$ 3 & VLT/NACO & \citet{Chauvin:2012} \\
2009-12-29 & 0.306 $\pm$ 0.009& 212.1 $\pm$ 1.7 & VLT/NACO & \citet{Chauvin:2012} \\
2010-04-10 & 0.346 $\pm$ 0.007& 209.9 $\pm$ 1.2 & VLT/NACO & \citet{Chauvin:2012} \\
2010-09-28 & 0.383 $\pm$ 0.011& 210.3 $\pm$ 1.7 & VLT/NACO & \citet{Chauvin:2012} \\
2010-11-16 & 0.387 $\pm$ 0.008& 212.4 $\pm$ 1.4 & VLT/NACO & \citet{Chauvin:2012} \\
2010-11-17 & 0.390 $\pm$ 0.013& 212 $\pm$ 2 & VLT/NACO & \citet{Chauvin:2012} \\
2011-02-01 & 0.408 $\pm$ 0.009& 211.1 $\pm$ 1.5 & VLT/NACO & \citet{Chauvin:2012} \\
2011-03-26 & 0.426 $\pm$ 0.013& 210.1 $\pm$ 1.8 & VLT/NACO & \citet{Chauvin:2012} \\
2009-12-03 & 0.339 $\pm$ 0.010 & 209.2 $\pm$ 1.7 & Gemini-South/NICI & \citet{Nielsen:2014} \\
2009-12-03 & 0.323 $\pm$ 0.010 & 209.3 $\pm$ 1.8 & Gemini-South/NICI & \citet{Nielsen:2014} \\
2010-12-25 & 0.407 $\pm$ 0.005 & 212.8 $\pm$ 1.4 & Gemini-South/NICI & \citet{Nielsen:2014} \\
2011-10-20 & 0.452 $\pm$ 0.003 & 211.6 $\pm$ 0.4 & Gemini-South/NICI & \citet{Nielsen:2014} \\
2011-10-20 & 0.455 $\pm$ 0.005 & 211.9 $\pm$ 0.6 & Gemini-South/NICI & \citet{Nielsen:2014} \\
2012-03-29 & 0.447 $\pm$ 0.003 & 210.8 $\pm$ 0.4 & Gemini-South/NICI & \citet{Nielsen:2014} \\
2012-03-29 & 0.448 $\pm$ 0.005 & 211.8 $\pm$ 0.6 & Gemini-South/NICI & \citet{Nielsen:2014} \\
2012-12-02 & 0.461 $\pm$ 0.014 & 211.9 $\pm$ 1.2 & Magellan/MagAO & \citet{Nielsen:2014} \\
2012-12-04 & 0.470 $\pm$ 0.010 & 212.0 $\pm$ 1.2 & Magellan/MagAO & \citet{Nielsen:2014} \\
2013-11-16 & 0.4308 $\pm$ 0.0015 & 212.43 $\pm$ 0.17 & Gemini-South/GPI & This Work\tablenotemark{1} \\
2013-11-16 & 0.4291 $\pm$ 0.0010 & 212.58 $\pm$ 0.15 & Gemini-South/GPI & This Work\tablenotemark{1} \\
2013-11-18 & 0.4302 $\pm$ 0.0010 & 212.46 $\pm$ 0.15 & Gemini-South/GPI & This Work\tablenotemark{1} \\
2013-12-10 & 0.4255 $\pm$ 0.0010 & 212.51 $\pm$ 0.15 & Gemini-South/GPI & This Work\tablenotemark{1} \\
2013-12-10 & 0.4244 $\pm$ 0.0010 & 212.85 $\pm$ 0.15 & Gemini-South/GPI & This Work\tablenotemark{1} \\
2013-12-11 & 0.4253 $\pm$ 0.0010 & 212.47 $\pm$ 0.16 & Gemini-South/GPI & This Work\tablenotemark{1} \\
2014-11-08 & 0.3562 $\pm$ 0.0010 & 213.02 $\pm$ 0.19 & Gemini-South/GPI & This Work\tablenotemark{1} \\
2015-04-02 & 0.3173 $\pm$ 0.0009 & 213.13 $\pm$ 0.20 & Gemini-South/GPI & This Work\tablenotemark{1} \\
2015-11-06 & 0.2505 $\pm$ 0.0015 & 214.14 $\pm$ 0.34 & Gemini-South/GPI & This Work\tablenotemark{1} \\
2015-12-05 & 0.2402 $\pm$ 0.0011 & 213.58 $\pm$ 0.34 & Gemini-South/GPI & This Work\tablenotemark{1} \\
2015-12-22 & 0.2345 $\pm$ 0.0010 & 213.81 $\pm$ 0.30 & Gemini-South/GPI & This Work\tablenotemark{1} \\
2016-01-21 & 0.2226 $\pm$ 0.0021 & 214.84 $\pm$ 0.44 & Gemini-South/GPI & This Work\tablenotemark{1} \\
2014-12-08 & 0.35051 $\pm$ 0.00320 & 212.60 $\pm$ 0.66 & VLT/SPHERE & \citet{lagrange:2018} \\
2015-05-05 & 0.33242 $\pm$ 0.00170 & 212.58 $\pm$ 0.35 & VLT/SPHERE & \citet{lagrange:2018} \\
2015-10-01 & 0.26202 $\pm$ 0.00178 & 213.02 $\pm$ 0.48 & VLT/SPHERE & \citet{lagrange:2018} \\
2015-11-30 & 0.24205 $\pm$ 0.00251 & 213.30 $\pm$ 0.74 & VLT/SPHERE & \citet{lagrange:2018} \\
2015-12-26 & 0.23484 $\pm$ 0.00180 & 213.79 $\pm$ 0.51 & VLT/SPHERE & \citet{lagrange:2018} \\
2016-01-20 & 0.22723 $\pm$ 0.00155 & 213.15 $\pm$ 0.46 & VLT/SPHERE & \citet{lagrange:2018} \\
2016-03-26 & 0.20366 $\pm$ 0.00142 & 213.90 $\pm$ 0.46 & VLT/SPHERE & \citet{lagrange:2018} \\
2016-04-16 & 0.19749 $\pm$ 0.00236 & 213.88 $\pm$ 0.83 & VLT/SPHERE & \citet{lagrange:2018} \\
2016-09-16 & 0.14236 $\pm$ 0.00234 & 214.62 $\pm$ 1.10 & VLT/SPHERE & \citet{lagrange:2018} \\
2016-10-14 & 0.13450 $\pm$ 0.00246 & 215.50 $\pm$ 1.22 & VLT/SPHERE & \citet{lagrange:2018} \\
2016-11-18 & 0.12712 $\pm$ 0.00644 & 215.80 $\pm$ 3.37 & VLT/SPHERE & \citet{lagrange:2018} \\
2018-09-17 & 0.14046 $\pm$ 0.00312 & 29.71 $\pm$ 1.67 & VLT/SPHERE & \citet{lagrange:2018} \\
2015-01-24 & 0.3355 $\pm$ 0.0009 & 212.88 $\pm$ 0.20 & Gemini-South/GPI & This Work \\
2018-09-21 & 0.1419 $\pm$ 0.0053 & 28.16 $\pm$ 1.82 & Gemini-South/GPI & This Work \\
2018-11-18 & 0.1645 $\pm$ 0.0018 & 28.64 $\pm$ 0.70 & Gemini-South/GPI & This Work \\
\enddata
\tablenotetext{1}{These epochs originally appeared in \citet{Wang:2016gl}, but have been recomputed here as a result of changes in the GPI pipeline, most noticeably the assumed North angle.}
\end{deluxetable*}

\subsection{Previously published datasets}

As in \citet{Nielsen:2014} and \citet{Wang:2016gl}, we compile relative astrometry of $\beta$ Pic b from the literature to extend the time baseline.  \citet{Chauvin:2012} presented nine epochs of data from VLT/NACO, including the two initial discovery epochs of 2003 \citep{Lagrange:2009hq} and 2009 \citep{Lagrange:2010}, up until 2011.  An additional seven epochs of data from 2009 to 2012 were reported from Gemini-South/NICI by  \citet{Nielsen:2014}, as well as two 2012 epochs from Magellan/MagAO \citep{morzinski:2015}.  Twelve epochs of Gemini-South/GPI data were presented by \citet{Wang:2016gl}, running from 2013 to 2016.  An additional attempt was made to observe $\beta$ Pic b with GPI on UT 2016-11-18, however given its proximity to the host star and the poor seeing that night, the planet was not detected in this dataset.  Recently, \citet{lagrange:2018} published eleven epochs of relative astrometry from VLT/SPHERE between 2014 and 2016, as well as an epoch from 2018-09-17 when the planet reappeared on the north-east side of the star.

Due to the timing issue and a change in the astrometric calibration \citep{derosa2019_north}, we also recomputed the astrometry of the epochs published in \citet{Wang:2016gl} using the same reduction parameters as the previous work. The parallactic angles in each dataset were recomputed with the correct time in the header following the procedure outlined in \citet{derosa2019_north}. We also used the new plate scale value of $14.161 \pm 0.021$~mas/pixel and varying residual North angle correction from \citet{derosa2019_north}. We used a residual North angle of $0.23^{\circ} \pm 0.11^{\circ}$ for the 2013 epochs, $0.17^{\circ} \pm 0.14^{\circ}$ for the 2014-11-08 and 2015-04-02 epochs, and $0.21^{\circ} \pm 0.23^{\circ}$ for the remaining 2015 and 2016 epochs. The recomputed astrometry is listed in Table \ref{tbl:imaging_data}. The most significant change to the astrometry presented here compared to \citet{Wang:2016gl} is the change in assumed North angle, from $-0.2^\circ$ to approximately $+0.2^\circ$, shifting all position angles to larger values by $\sim$0.4$^\circ$.  Additionally, we include an additional epoch from 2015-01-24, which had been initially rejected in \citet{Wang:2016gl} due to artefacts at the location of the planet.  With the rereduction, the artefacts are no longer visible, and we include this epoch in our final dataset.

\section{Orbit fitting}

\subsection{Hipparcos Intermediate Astrometric Data}

The \textit{Hipparcos} mission performed detailed astrometric monitoring of bright stars, with the majority of stars (including $\beta$ Pic) being fit by a five-parameter solution, RA and Dec of the star (as would be observed from the Solar System barycenter) at a reference epoch of 1991.25, parallax, and proper motion in RA and Dec.  Individual measurements were made of each star along a one-dimensional scan referred to as the abscissa, with no published constraints in the direction perpendicular to the scan direction.  The direction of the scan changed from orbit to orbit as the satellite surveyed the sky, allowing a two dimensional motion to be reconstructed from a series of one-dimensional measurements.  \citet{vanLeeuwen:2007dc} provides Intermediate Astrometric Data (IAD) from the rereduction of the \textit{Hipparcos} data in the form of a DVD-ROM attached to the book, which include scan directions, residuals from the fit, and errors on the measurement, for each epoch of data.

While the IAD do not contain the abscissa measurements themselves, the measurements can be reconstructed from these values.  We extract from the \citet{vanLeeuwen:2007book} IAD the epoch of the orbit in decimal years ($t$), scan direction ($\sin(\phi)$ and $\cos(\phi)$), residual to the best fit ($R$), and error on the original measurement ($\epsilon$).  This is combined with the best fitting solution from the \citet{vanLeeuwen:2007dc} catalog for the star, which provides the five astrometric parameters, $\alpha_{0}$, $\delta_{0}$, $\pi$, $\mu_{\alpha^*}$, $\mu_{\delta}$: the right ascension and declination at the \textit{Hipparcos} reference epoch of 1991.25 in degrees, the parallax in mas, and the proper motion in right ascension and declination in mas/yr.  The notation $\mu_{\alpha^*}$ indicates offsets and velocities in right ascension are multiplied by $\cos \delta_0$, to prevent a constant factor between the magnitude of offsets in right ascension and declination.

We first find the ephemeris for the star over the epochs of \textit{Hipparcos} measurements ($t$) from the best-fit astrometric parameters:

\begin{equation}
\begin{split}
\Delta \alpha^* (t) = \pi (X(t) \sin(\alpha_{0}) - Y(t) \cos(\alpha_{0})) \\
+ (t - 1991.25) \mu_\alpha^* 
\end{split}
\end{equation}

\noindent and 

\begin{equation}
\begin{split}
    \Delta \delta (t) = \pi (X(t) \cos(\alpha_{0}) \sin(\delta_{0}) \\
    + Y(t) \sin(\alpha_{0}) \sin(\delta_{0}) - Z(t) \cos(\delta_{0})) \\
    + (t - 1991.25) \mu_\delta
\end{split}
\end{equation}

\noindent $\Delta \alpha^* (t)$ and $\Delta \delta (t)$ represent the offset from the catalog position ($\alpha_0$, $\delta_0$) at the solar system barycenter of the photocenter from proper motion and parallax only.  $X$, $Y$, and $Z$ in au are the location of the Earth in barycentric coordinates.  With this ephemeris, we can then reconstruct the abscissa measurement for each \textit{Hipparcos} epoch.  The residual gives the difference between this ephemeris and the \textit{Hipparcos} measurement at a time $t$, along the scan direction $\phi$.  The abscissa measurement, then, is a line that passes through the point:

\begin{equation}
\begin{split}
    \alpha_a^*(t) = R(t) \cos(\phi(t)) + \Delta \alpha^*(t) \\
    \delta_a(t) = R(t) \sin(\phi(t)) + \Delta \delta(t) 
\end{split}
\end{equation}

\noindent where for convenience, $\alpha_{a}^*$ and $\delta_{a}$ are offsets from (0,0), taken to be the \textit{Hipparcos} catalog values of $\alpha_{0}$ and $\delta_{0}$.  The \textit{Hipparcos} measurement is one-dimensional, and so consists of a line through the point $(\alpha_a^*(t),\delta_a(t))$, but perpendicular to the scan direction.  We define such a line by two points each separated by 1~mas from  $(\alpha_a^*(t),\delta_a(t))$, 

\begin{equation}
\begin{split}
    \alpha_M^*(t) = [-1,1] \times \sin(\phi(t)) + \alpha_a^*(t) \\
    \delta_M(t) = [1,-1] \times \cos(\phi(t)) + \delta_a(t)
\end{split}
\end{equation}

\noindent So the Hipparcos measurement at epoch $t$ is then given by a line passing through the points defined by $\alpha_M^*(t)$ and $\delta_M(t)$.  The error from \citet{vanLeeuwen:2007book} ($\epsilon$) is the distance in mas from this line in the perpendicular direction (along the scan direction).  These measurements and errors can then be fit with any astrometric model, either the 5-parameter fit performed by \citet{vanLeeuwen:2007dc}, or a more complicated combination of these parameters and orbital parameters.  For arbitrary functions that give calculated values of position as a function of time $\alpha_C^*(t)$ and $\delta_C(t)$, $\chi^2$ can be calculated by first finding the residual separation ($d$) from the measurement in the perpendicular direction (along the \textit{Hipparcos} scan direction), using the equation for the distance from a point $(x_0,y_0)$ to a line defined by the points $(x_1,y_1)$ and $(x_2,y_2)$:

\begin{equation}
    d = \frac{|(y_2 - y_1)x_0 - (x_2 - x_1)y_0 + x_2y_1 - y_2x_1|}{\sqrt{(y_2-y_1)^2 + (x_2-x_1)^2}}
\end{equation}

\noindent when we substitute ($x_1$, $x_2$) = $\alpha_M^*(t)$, ($y_1$, $y_2$) = $\delta_M(t)$, $x_0 = \alpha_C^*(t)$, and $y_0 = \delta_C(t)$ the expression for $d$ simplifies to:

\begin{equation}
\begin{split}
d(t) = |(\alpha_a^*(t) - \alpha_C^*(t)) \cos(\phi(t)) \\
+ (\delta_a(t) - \delta_C(t)) \sin(\phi(t))|
\end{split}
\end{equation}

\noindent which allows us to calculate the $\chi^2$ of a given model from

\begin{equation}
    \chi^2 = \sum_{t} \left ( \frac{d(t)}{\epsilon(t)} \right)^2
\label{eq:chisq}
\end{equation}

To test the consistency of this method, we extract the abscissa measurements of $\beta$ Pic from \citet{vanLeeuwen:2007dc} and \citet{vanLeeuwen:2007book}, which consist of 111 epochs between 1990.005 and 1993.096, and then refit them with the same 5-parameter model.  We use a Metropolis-Hastings MCMC procedure \citep{Nielsen:2014} to sample the posterior of the five parameters $\alpha_{H0}^*$, $\delta_{H0}$, $\pi$, $\mu_{\alpha^*}$, $\mu_{\delta}$, and compare to the values and errors given by \citet{vanLeeuwen:2007dc}.  We define $\alpha_{H0}^*$ and $\delta_{H0}$ as the offsets in mas of the photocenter in 1991.25 from the \textit{Hipparcos} catalog positions $\alpha_0$ and $\delta_0$ as measured from the solar system barycenter.  Thus, in this five-parameter fit our model has values for $\alpha_C^*(t)$ and $\delta_C(t)$ of

\begin{equation}
\begin{split}
    \alpha_C^*(t) = \alpha_{H0}^* + \pi (X(t) \sin(\alpha_{0}) - Y(t) \cos(\alpha_{0})) \\
    + (t - 1991.25) \mu_\alpha^*
\end{split}
\end{equation}

\noindent and 

\begin{equation}
\begin{split}
    \delta_C(t) = \delta_{H0} + \pi (X(t) \cos(\alpha_{0}) \sin(\delta_{0}) \\
    + Y(t) \sin(\alpha_{0}) \sin(\delta_{0}) - Z(t) \cos(\delta_{0}))  \\
    + (t - 1991.25) \mu_\delta
\end{split}
\end{equation}

\begin{figure}[ht]
\includegraphics[width=\columnwidth,trim=2cm 8cm 2cm 7cm]{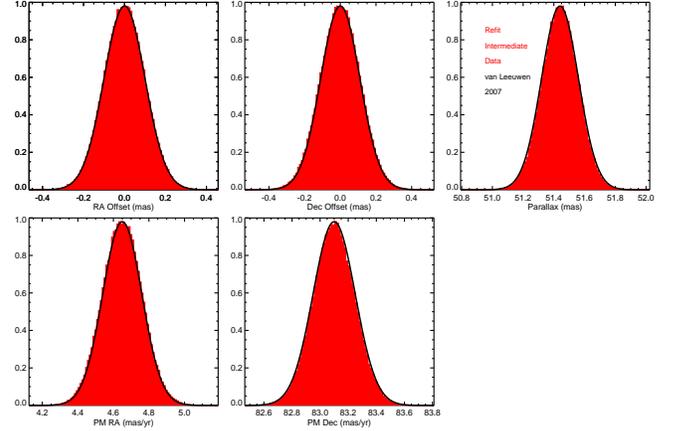}
\caption{Refit of extracted \textit{Hipparcos} IAD abscissa measurements, with the 1D posterior on each parameters from our MCMC fit to the IAD shown in the red filled histogram, and a Gaussian probability distribution using the \textit{Hipparcos} catalog values and errors for each parameter shown as an overplotted black curve. We find excellent agreement between our MCMC fit and the \citet{vanLeeuwen:2007dc} \textit{Hipparcos} catalog values. \label{fig:hipfig1}}
\end{figure}

A simple fit of the extracted abscissa values and errors produces posteriors with median values that match the catalog values, but with standard deviations that are $\sim$10\% too large.  This discrepancy arises because the catalog errors are renormalized to achieve $\chi_\nu^2 = 1$;  to reproduce this renormalization, we multiply the individual errors on each abscissa measurement ($\epsilon(t)$) by a factor $f$:
\begin{equation}
    f = D \left ( G \sqrt{ \frac{2}{9 D}}  + 1 - \left (\frac{2}{9 D} \right ) \right )^3
\end{equation}
\noindent where $D$ is the number of degrees of freedom ($N_{\rm data} - N_{\rm parameters} - 1 = N_{\rm epochs} - 6$) and G is the goodness of fit \citep{michalik:2014}.  The value for $G$ for $\beta$ Pic is $-1.63$, as given by \citet{vanLeeuwen:2007dc}.   Figure~\ref{fig:hipfig1} shows the comparison after performing this renormalization of the errors, with our fit in the filled red histogram, and the \citet{vanLeeuwen:2007dc} \textit{Hipparcos} catalog values represented as the black curve, taken to be a Gaussian with mean equal to the catalog measurement, and standard deviation the catalog error.  The two match to within the numerical precision of the catalog values.  We conclude that the abscissa measurements we extract from the \textit{Hipparcos} IAD are suitable for including in our orbit fits of the system.

\subsection{Gaia DR2}
The \textit{Gaia} DR2 magnitude of \target\ is $G=3.72$ and it is therefore a star that lies outside the nominal magnitude range of the \textit{Gaia} mission \citep{GaiaCollaboration:2016aa}. It is being observed because small improvements to the onboard detection parameters were made before routine operations began \citep{Sahlmann:2016ac, Martin-Fleitas:2014aa}. However, it can be expected that the degraded astrometric performance for bright stars in the range $G=5-6$ observed in DR2 \citep[e.g.][Fig. 9]{Lindegren:2018aa} is even more pronounced for brighter stars like \target. The data for \target\ in Gaia DR2 have therefore to be treated with additional caution.

\begin{figure}
\gridline{\fig{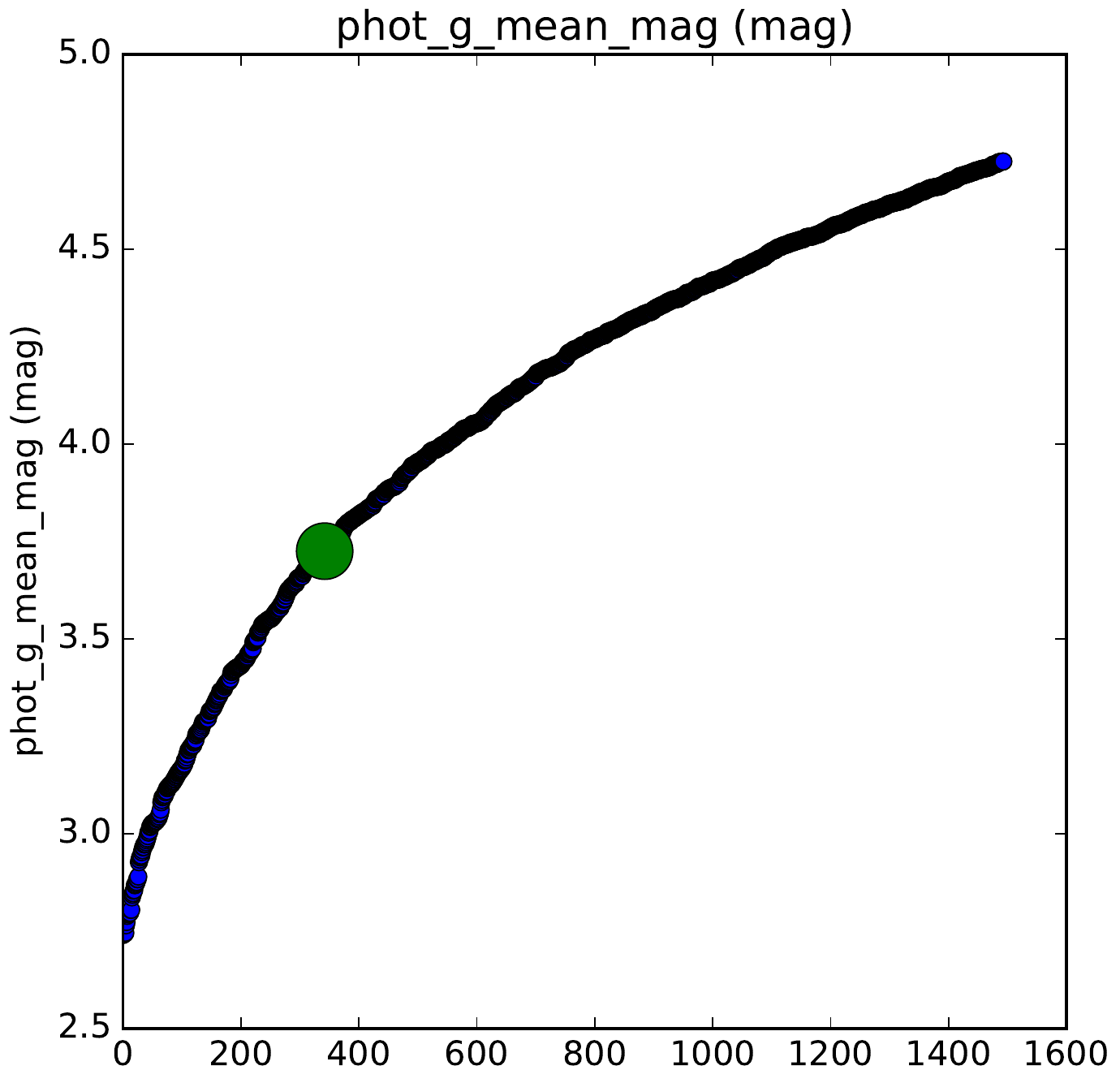}{0.5\columnwidth}{(a)}
\fig{{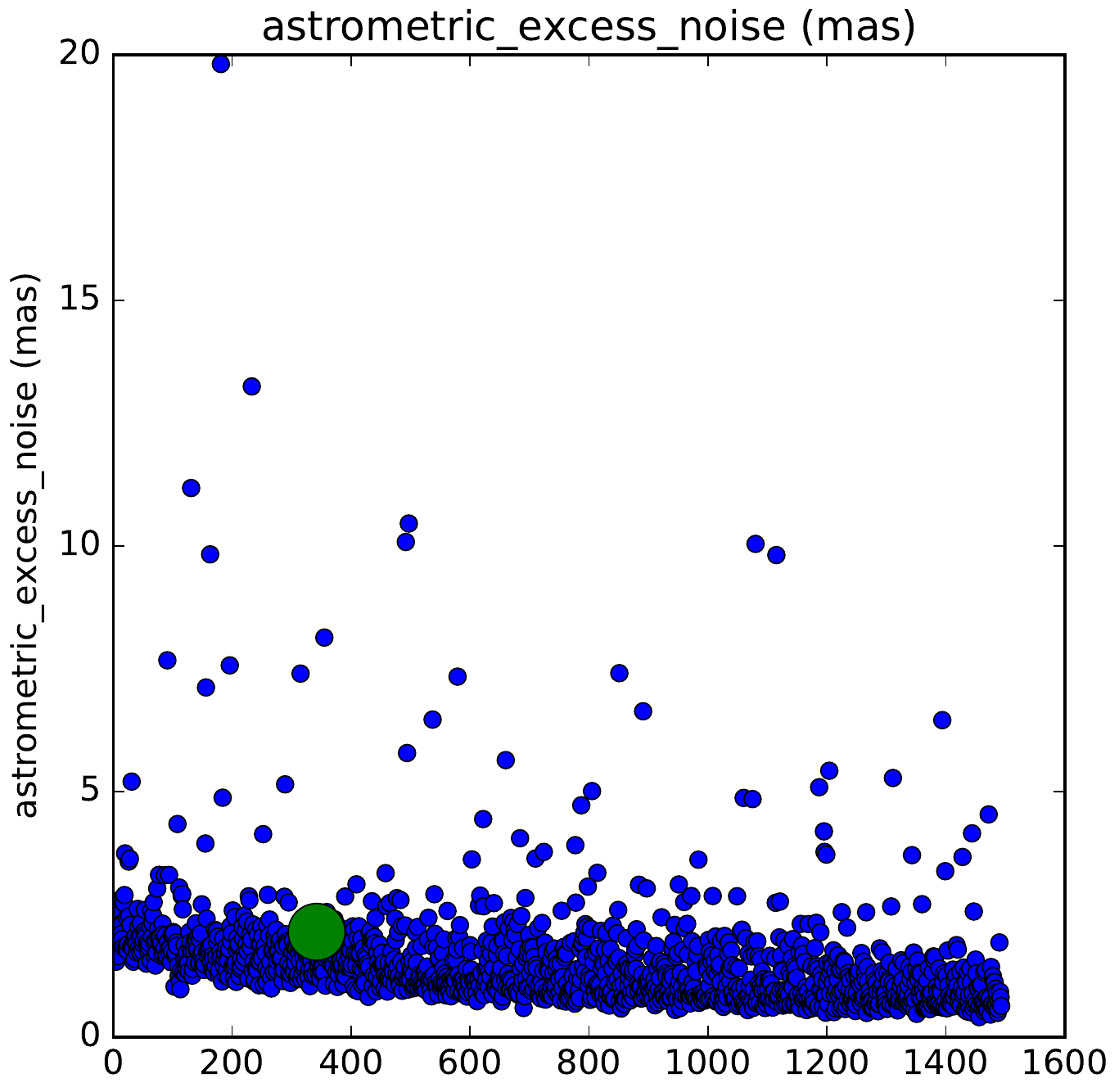}}{0.5\columnwidth}{(b)}
          }
\gridline{\fig{{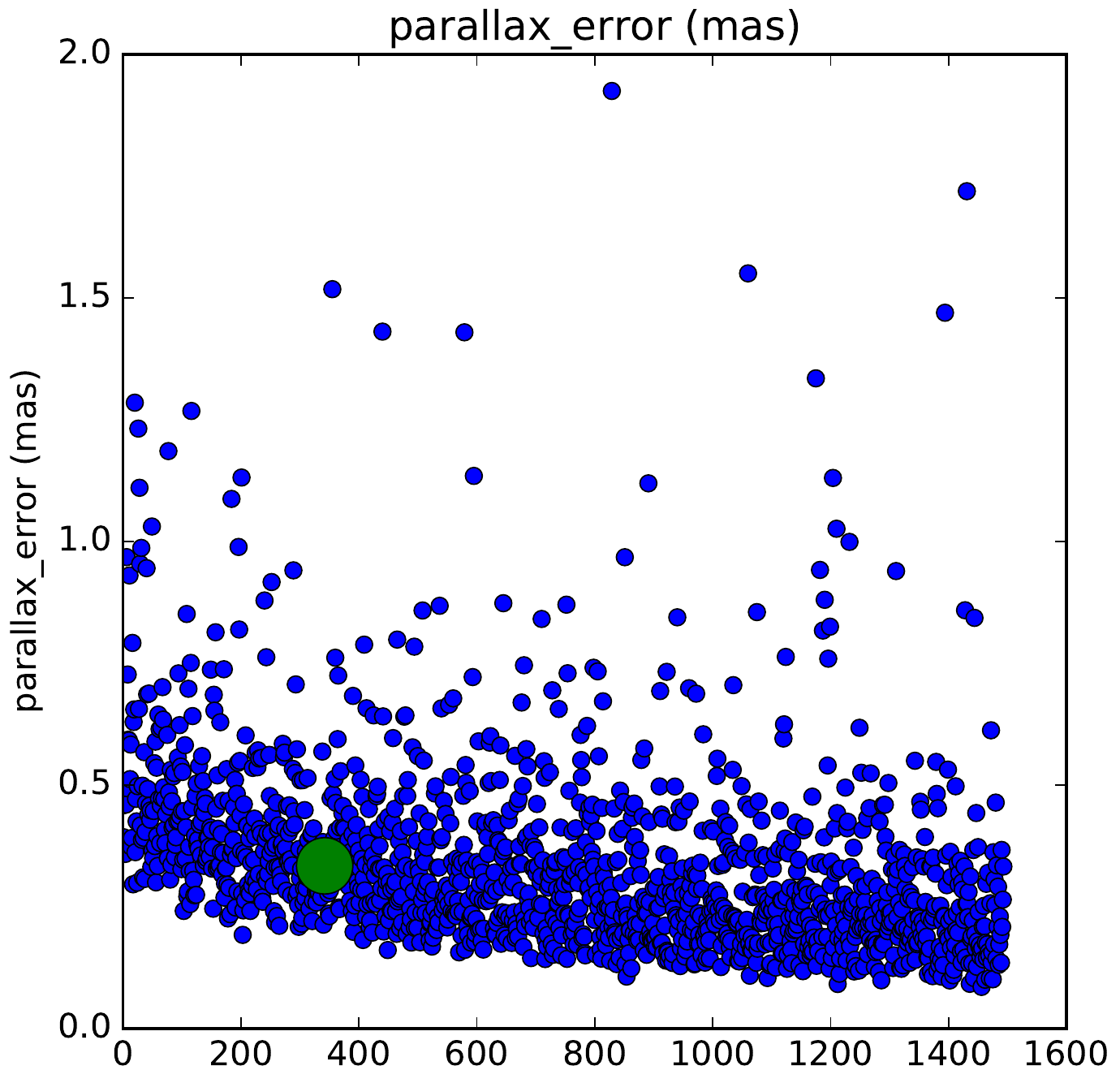}}{0.5\columnwidth}{(c)}
\fig{{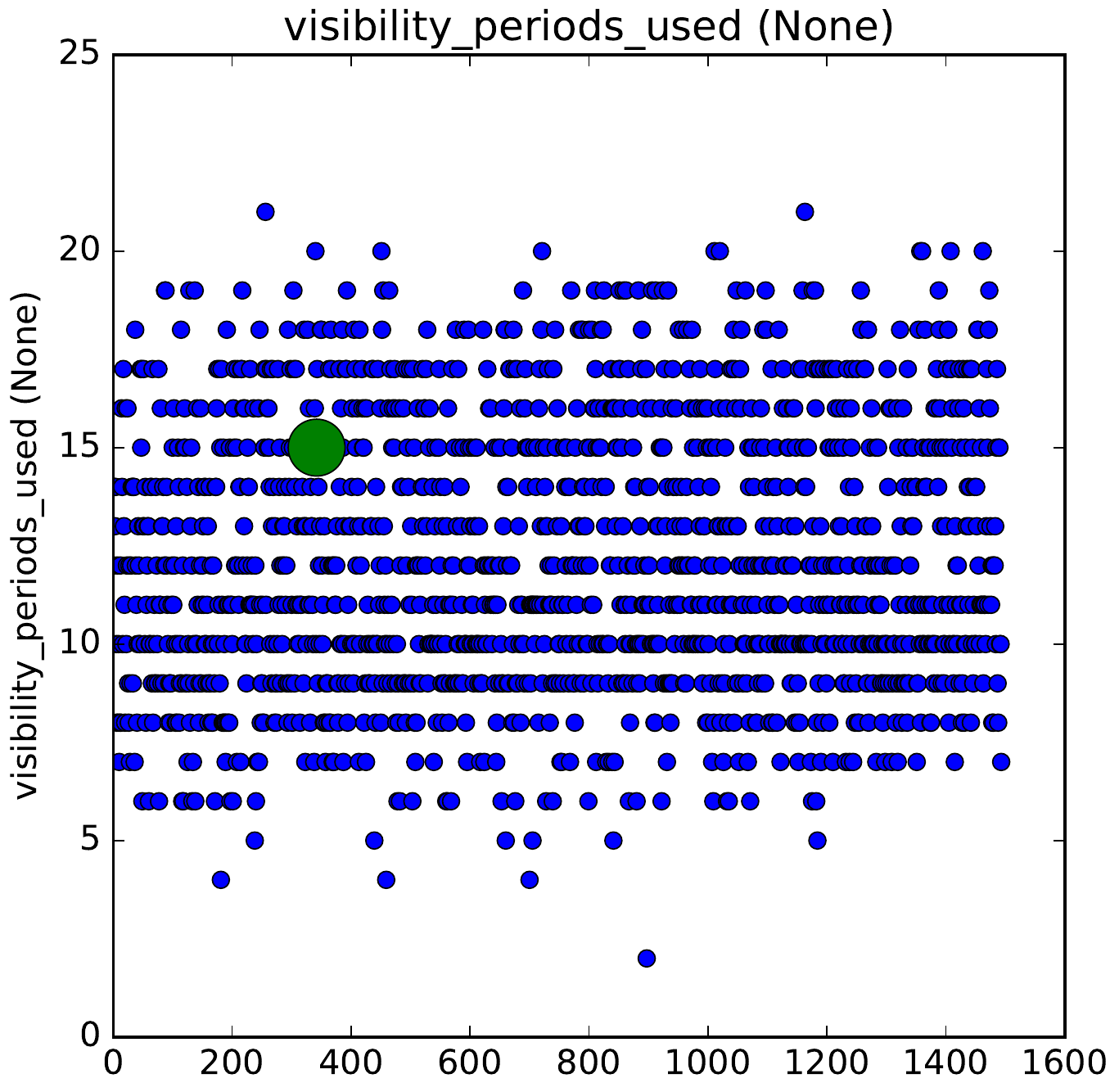}}{0.5\columnwidth}{(d)}
          }
\caption{Gaia DR2 parameters of \target\ (large green circle) compared with $\sim$1500 stars with similar magnitudes. The x-axis is the star sequence number.\label{fig:comparison}}
\end{figure}

To establish a notion of the quality of the DR2 data, we compared several quality indicators for a comparison sample of stars, chosen to have magnitudes within $\pm1$ of \target. We used \href{https://github.com/Johannes-Sahlmann/pygacs}{pygacs}\footnote{\url{https://github.com/Johannes-Sahlmann/pygacs}} to query the Gaia archive and retrieved 1494 very bright stars with $G=2.72-4.72$. Figure \ref{fig:comparison} shows a small selection of DR2 catalog parameters and we inspected many more. From this comparison, \target\ appears to be a `typical' very bright star in terms of excess noise, parameter uncertainties, and number of Gaia observations, with no indication of being particularly problematic. 

In particular, the \texttt{astrometric\_excess\_noise} of 2.14 mas is large when compared to stars in the nominal Gaia magnitude range, but not outstanding when compared to other very bright stars. If the excess noise would be normally distributed, we expect it to average out with $1/\sqrt{\mathtt{astrometric\_matched\_observations}=30}$, yielding 0.39 mas which is comparable with the DR2 errors in positions and parallax (0.32 -- 0.34 mas).

As in \citet{snellen:2018}, we also make use of the \textit{Gaia} DR2 data \citep{GaiaCollaboration:2018io} to further constrain the orbit.  The Intermediate Astrometric Data from \textit{Gaia} are not yet publicly available, and so we can only utilize the catalog values from the 5-parameter fit.  As \citet{snellen:2018} note,  $\alpha_G$ and $\delta_G$, the solar system barycentric coordinates of the star at \textit{Gaia} reference epoch of 2015.5 strongly constrains the proper motion, given the long time baseline to the 1989-1993 \textit{Hipparcos} data.  As both measurements are in the solar system barycentric frame (ICRS J2000), the offset between ($\alpha_G$, $\delta_G$) and the Hipparcos values ($\alpha_0$, $\delta_0$) at the reference epoch of 1991.25 should be a combination of proper motion of the system and orbital motion.

\subsection{Orbit Fitting Results}

\subsubsection{Relative astrometry only}
\begin{figure*}[ht]
\includegraphics[width=\textwidth,trim=2cm 3.5cm 2cm 9cm]{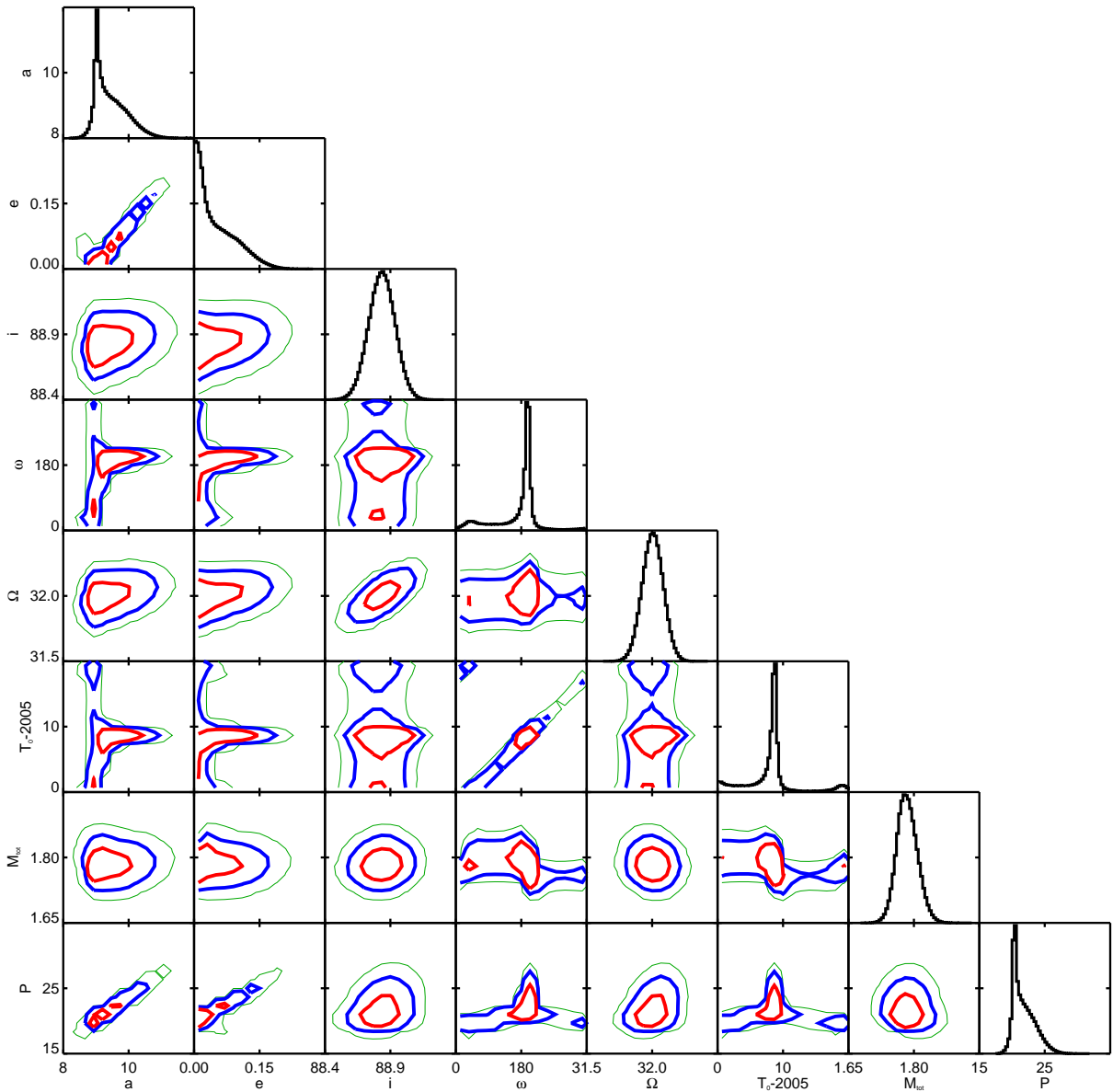}
\caption{Triangle plot for the orbit fit to $\beta$ Pic b using only the imaging data from NaCo, NICI, Magellan, and GPI (Case 2).  A strong degeneracy exists between eccentricity and semi-major axis, with more eccentric orbits having longer periods. \label{fig:nohip_triangle}}
\end{figure*}

\begin{figure*}[ht]
\includegraphics[width=\textwidth,trim=2cm 10.5cm 2cm 8cm]{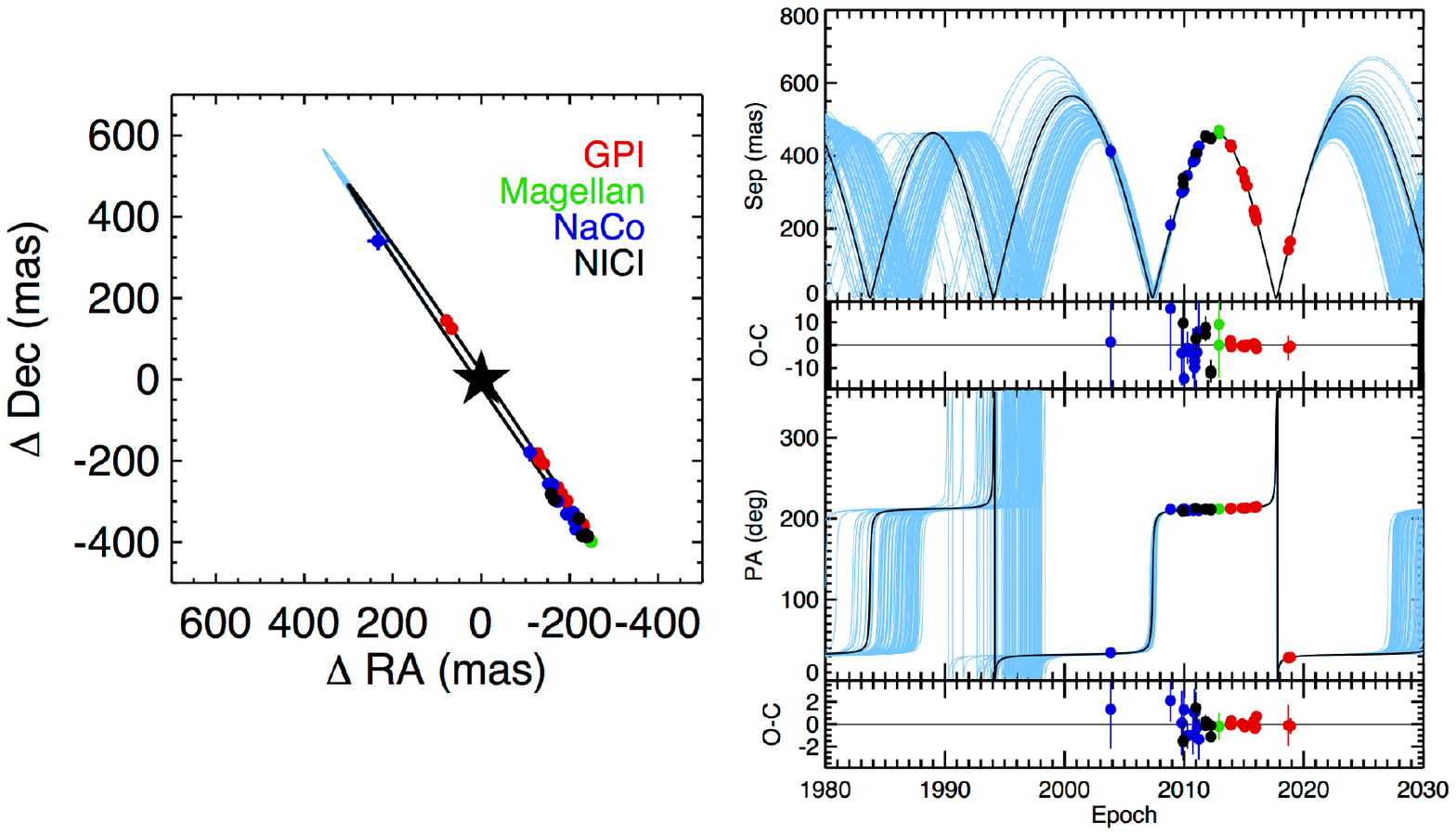}
\caption{Orbit tracks for the orbit fit using only the imaging data (Case 2).  The black line shows the lowest $\chi^2$ orbit, while the blue curves are 100 sets of orbital parameters drawn from the posterior. \label{fig:nohip_tracks}}
\end{figure*}

Before including the \textit{Hipparcos} and \textit{Gaia} data, we begin by fitting an orbit to the direct imaging data alone.  We again utilize the MCMC Metropolis Hastings orbit fitting procedure described previously in \citet{Nielsen:2014}, \citet{Nielsen:2016ct}, and \citet{nielsen:2017}.  We perform a fit in seven parameters, with the typical priors for visual orbits, semi-major axis ($a$) uniform in log($a$) ($\frac{dN}{d\log{a}} \propto C$, which is equivalent to $\frac{dN}{da} \propto a^{-1}$), uniform eccentricity ($e$), inclination angle ($i$) uniform in $\cos(i)$, and uniform in argument of periastron ($\omega$), position angle of nodes ($\Omega$), epoch of periastron passage ($T_0$), and total mass ($M_{tot}$).  Period ($P$) is then derived from $a$ and $M_{tot}$ using Kepler's third law.  The distance for this fit is set to be fixed at the \textit{Hipparcos} value of 19.44 pc \citep{vanLeeuwen:2007dc}.  To avoid systematic offets between different instruments as much as possible, we do not fit the SPHERE data from \citet{lagrange:2018}, and limit our fit to the dataset of \citet{Wang:2016gl}.  Fits to imaging datasets have a well-known degeneracy in orbital parameters between [$\omega$,$\Omega$] and [$\omega$+180$^\circ$,$\Omega$+180$^\circ$], a degeneracy that is classically broken  with RV observations.  In the case of $\beta$ Pic b, a radial velocity measurement has been made for the planet itself by \citet{snellen:2014}, who find the RV of the planet, with respect to the host star, to be $-15.4 \pm 1.7$ km/s, at 2013-12-17.  We include this RV datapoint in this and subsequent fits.  We refer to this dataset as ``Case 1.''  Given the changes to the GPI astrometry, we find an orbit fit that is shifted toward lower periods and more circular orbits.  \citet{Wang:2016gl} reported [$a$, $e$, $i$, $\omega$, $\Omega$, $\tau$, P, M$_{tot}$] of [9.66$^{+1.12}_{-0.64}$ au, 0.080$^{+0.091}_{-0.053}$, 88.81${^{+0.12}_{-0.11}}^\circ$, 205.8${^{+52.6}_{-13.0}}^\circ$, 31.76${^{+0.80}_{-0.09}}^\circ$, 0.73$^{+0.14}_{-0.41}$, 22.47$^{+3.77}_{-2.26}$ yrs, 1.80$^{+0.03}_{-0.04}$M$_{\odot}$], compared to our values for "Case 1" of [8.95$^{+0.30}_{-0.32}$ au, 0.0360$^{+0.029}_{-0.022}$, 88.80$\pm 12 ^\circ$, 290.8${^{+60.0}_{-73.8}}^\circ$, 32.02$\pm 0.09^\circ$, 1.14$^{+0.22}_{-0.26}$, 20.18$^{+1.05}_{-0.97}$ yrs, 1.75$\pm 0.03$M$_{\odot}$].  \citet{Wang:2016gl} define epoch of periastron passage, $\tau$, as the number of orbital periods from MJD=50000 (1995.7726), and we converted our value of T$_0$ to this convention for this comparison.

Next, we repeat the orbit fit, but including the two additional epochs of GPI data from 2018 described in Section~\ref{sec:gpi_data}, which we refer to as ``Case 2.''  We display the posteriors for this fit in Figure~\ref{fig:nohip_triangle}.  The orbits themselves are shown in Figure~\ref{fig:nohip_tracks}, with posteriors for this and all other orbits given in Table~\ref{tbl:posteriors}.

Generally, low eccentricity orbits are preferred, with a peak at $e$=0, with a strong correlation between eccentricity and semi-major axis.  Periastron is preferred to be near 2014 (2013.5$^{+3.4}_{-0.7}$), with non-zero probability across 20 years, corresponding to circular orbits where periastron is undefined.

\begin{figure*}[ht]
\includegraphics[width=\textwidth,trim=2cm 3.5cm 2cm 21cm]{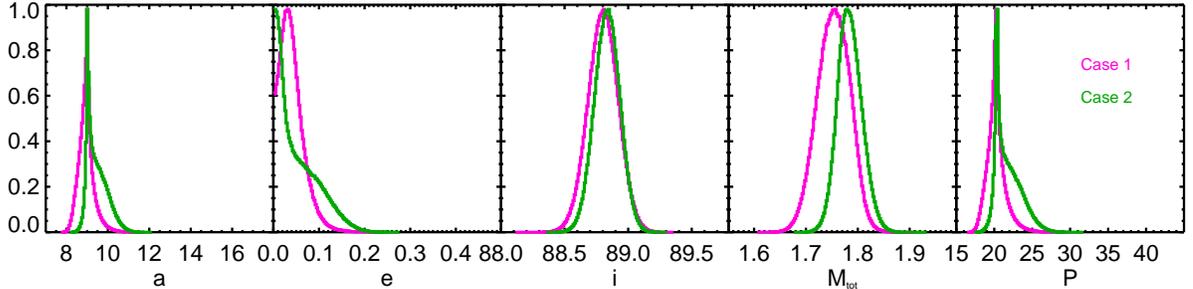}
\caption{Posterior probability distributions for semi-major axis (au), eccentricity, inclination angle (degrees), total mass (M$_\odot$), and period (yrs), for the imaging-only fit with the new 2018 GPI data (Case 2, green) and without (Case 1, pink).  The new data, following conjunction, result in more probability at orbits with larger eccentricity and semi-major axis. \label{fig:gpipoint}}
\end{figure*}

Figure~\ref{fig:gpipoint} compares posteriors on five parameters for the Case 1 and Case 2 fits.  Including GPI data after conjunction results in higher probability of more eccentric orbits, larger periods, and larger total mass for the system.

\subsubsection{Relative and absolute astrometry}
We next include the \textit{Hipparcos} and \textit{Gaia} data in our fit.  In addition to the previous seven parameters ($a$, $e$, $i$, $\omega$, $\Omega$, $T_0$, $M_{tot}$), we add six more for a total of thirteen.  The additional parameters are mass of the planet in M$_{Jup}$ ($M_P$), the location of the star from the Solar System  barycenter at the \textit{Hipparcos} reference epoch of 1991.25 ($\alpha_{H0}^*$, $\delta_{H0}$, both expressed as an offset from the \citealt{vanLeeuwen:2007dc} catalog positions, in mas), parallax ($\pi$) in mas, and proper motion ($\mu_{\alpha^*}$, $\mu_\delta$), in mas/yr.  As before, $\alpha_{H0}^*$ and $\mu_{\alpha^*}$ indicate $\alpha_{H0} \cos(\delta_0)$ and $\mu_\alpha \cos(\delta_0)$, in order to correct for the non-rectilinear nature of the coordinate system.  Uniform priors are assumed for all six additional parameters.

Our dataset includes the imaging data and planet RV used in the previous fit, as well as our extracted abscissa measurements and errors from the \textit{Hipparcos} IAD, and the \textit{Gaia} DR2 values of $\alpha_G$ and $\delta_G$ and associated errors.  $\chi^2$ then has four components.  The first is the standard separation and position angles for the imaging data and errors, with calculated values taken from the seven imaging data orbital parameters, and the distance taken from the parallax parameter.  The second is the CRIRES radial velocity of the planet from \citet{snellen:2014}, with reported errors.

The third component, the IAD contribution, comes from Equation~\ref{eq:chisq} and all thirteen parameters, with the position as a function of \textit{Hipparcos} epoch calculated from the standard five astrometric parameters, and additional displacement given by the motion of $\beta$ Pic around the center of mass of the star/planet system.  We approximate $\beta$ Pic b as having zero flux in the \textit{Hipparcos} and \textit{Gaia} bandpasses.  From the BT-Settl models  \citep{baraffe:2015}, at 26 Myr and 20 M$_{Jup}$ (well above the expected mass of $\sim$12 M$_{Jup}$), $\beta$ Pic b would have an apparent magnitude in the \textit{Gaia} $G$ bandpass of 16.9 mags, 13.1 mags fainter than $\beta$ Pic.  From our MCMC fit to the visual data alone, the maximum value of apastron reached was 0.8"; even at this value the offset between the photocenter and the star itself in the \textit{Gaia} data is 0.005 mas, well below the precision of any of the measurements.  The parameters for the visual orbit give the motion of the planet around the star ($\Delta \alpha_V^*$, $\Delta \delta_V$), and so the motion of the star around the barycenter is then $\Delta \alpha_s^* = - \Delta \alpha_V^* \frac{M_p}{M_{tot}}$, and similarly for $\Delta \delta_s$.  The value of $\Delta \alpha_s^*$ and $\Delta \delta_s$ are calculated at 1991.25 and subtracted from each \textit{Hipparcos} epoch to give the relative motion since the reference values of $\alpha_0$ and $\delta_0$.

The final components of the $\chi^2$ come from fits to the \textit{Gaia} values of $\alpha_G$ and $\delta_G$.  We fit the offset between these two values and our fit parameters, ($\alpha_G - \alpha_0 - \alpha_{H0}^*$) $\times$ $\cos{\delta_0}$ and $\delta_G - \delta_0 - \delta_{H0}$, with errors given by the stated errors in the \textit{Gaia} DR2 catalog.  We then fit this offset from the combination of the astrometric motion from $\mu_{\alpha^*}$ and $\mu_\delta$, as well as the orbital motion using the same method as for the \textit{Hipparcos} IAD.  We do not incorporate corrections to the non-rectilinear coordinate system or relativistic effects described by \citet{butkevich:2014}, given the \textit{Gaia} error bars are significantly larger than the magnitude of these effects.

\begin{figure*}[ht]
\includegraphics[width=\textwidth,trim=2cm 3.5cm 2cm 8cm]{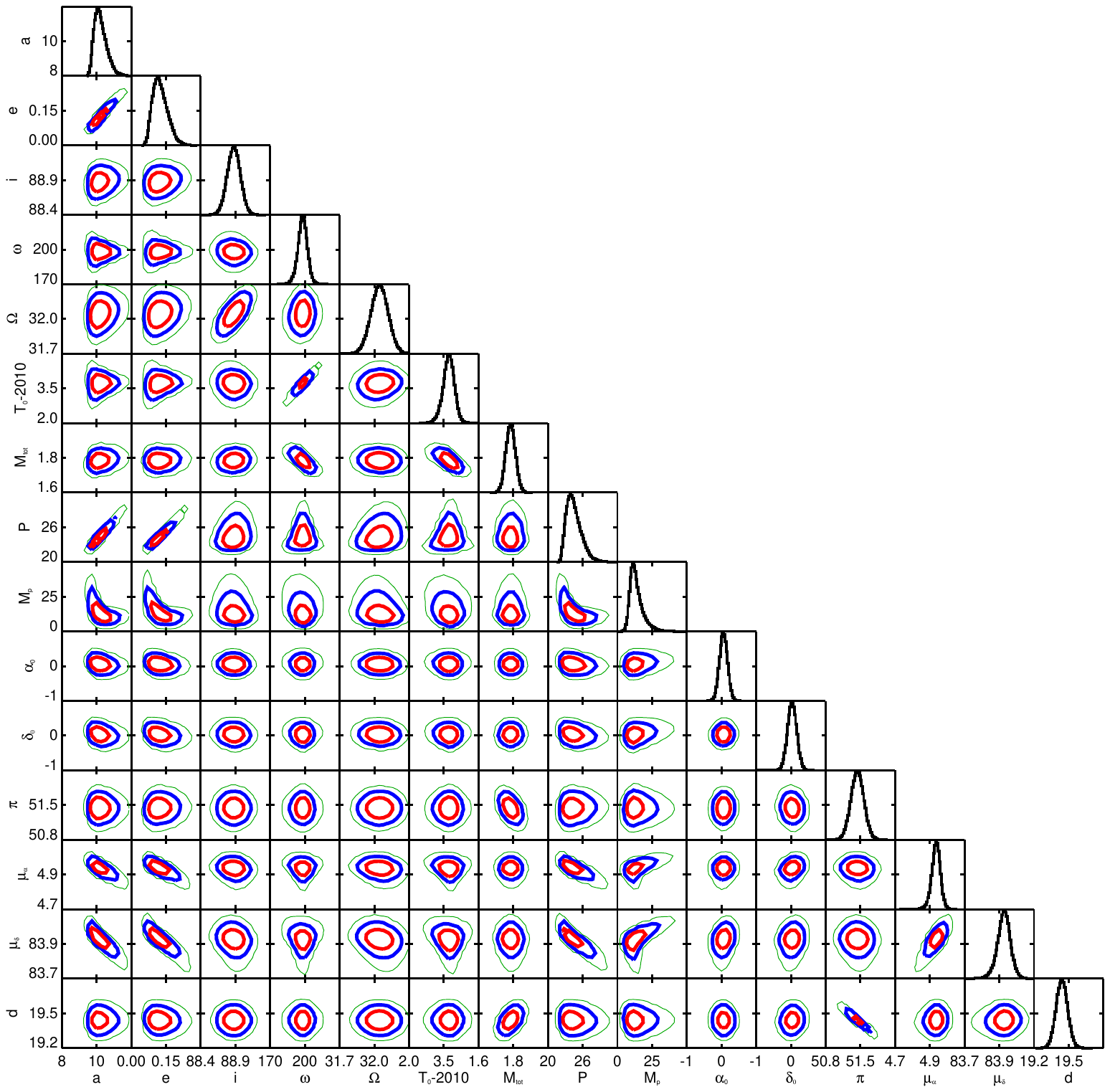}
\caption{Triangle plot for the orbit fit to the imaging dataset of NaCo, NICI, MagAO, and GPI, the CRIRES RV, as well as the astrometric data from \textit{Hipparcos} and \textit{Gaia} (Case 3).  With the addition of the astrometry, slightly larger eccentricities are preferred, and thus slightly larger orbital periods. \label{fig:triangle}}
\end{figure*}

\begin{figure*}[ht]
\includegraphics[width=\textwidth,trim=2cm 10cm 2cm 8cm]{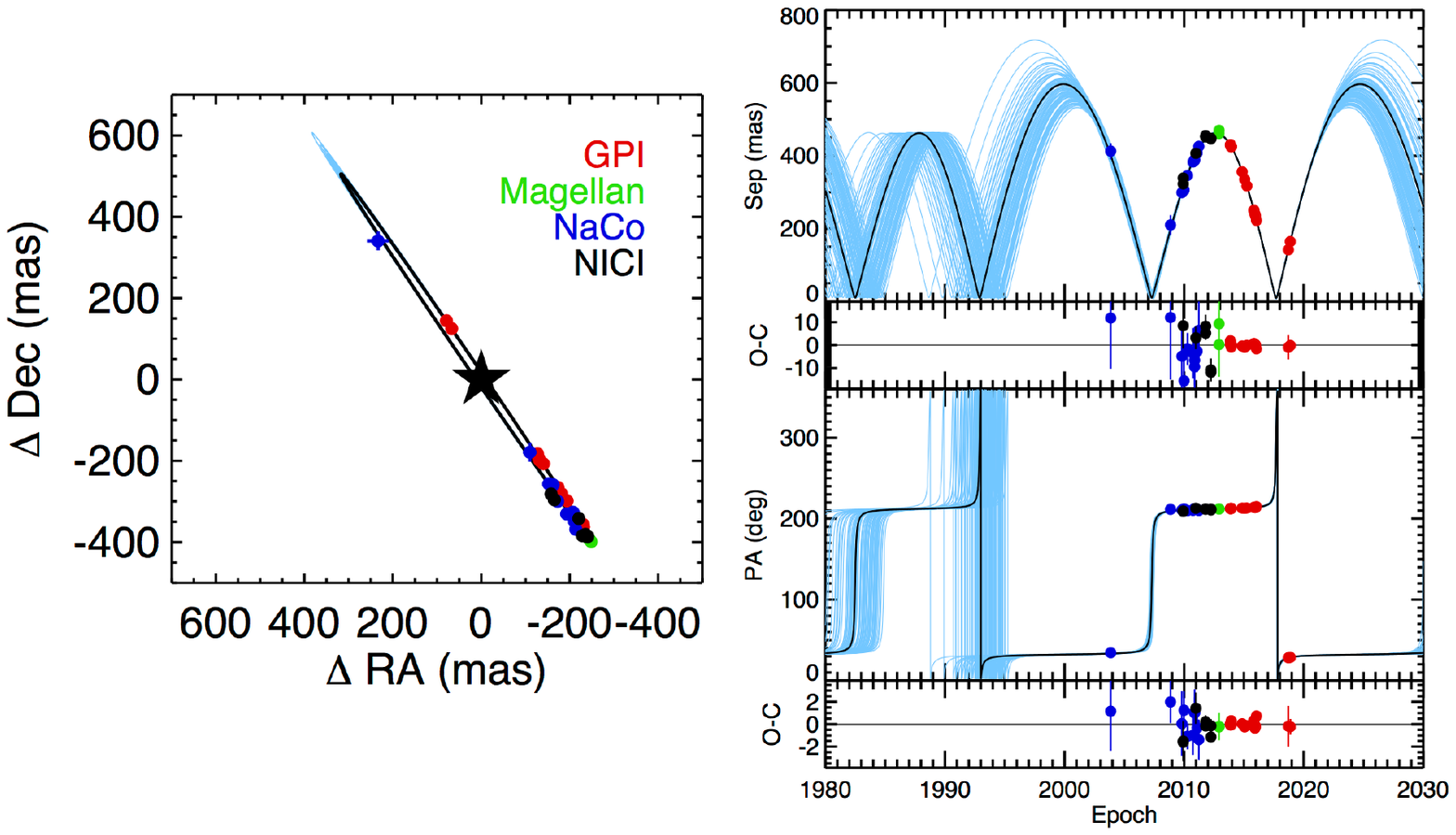}
\caption{Orbit tracks for the orbit fit to both the imaging and astrometric datasets (Case 3).  While the uncertainty in eccentricity remains, with zero eccentricity orbits no longer allowed, longer orbital periods are preferred. \label{fig:tracks}}
\end{figure*}

\begin{deluxetable*}{cccc}
\tablewidth{0pt}
\tabletypesize{\small}
\tablecaption{Properties of the $\beta$ Pic system\label{tbl:system}}
\tablehead{ \colhead{  } & \colhead{$\beta$ Pic} & \colhead{$\beta$ Pic b} & \colhead{Ref.}}
\startdata
\hline
$\alpha$ (deg) & 86.82123366090 &  &  \citet{GaiaCollaboration:2018io} \\
$\delta$ (deg) & -51.06614803159 &  & \citet{GaiaCollaboration:2018io} \\
$\mu_{\alpha^*}$ (mas/yr) & \multicolumn{2}{c}{4.94$\pm 0.02$} & this work \\
$\mu_\delta$ (mas/yr) & \multicolumn{2}{c}{83.93$^{+0.03}_{-0.04}$} & this work \\
$\pi$ (mas) & \multicolumn{2}{c}{51.44$\pm$0.13} & this work \\
d (pc) & \multicolumn{2}{c}{19.44$\pm$0.05} & this work \\
\hline
M & 1.77$\pm$0.03 M$_\odot$ & 12.8$^{+5.5}_{-3.2}$ M$_{\rm Jup}$ & this work \\
$\log{\frac{L}{L_\odot}}$ & & -3.76$\pm$0.02 & \citet{Chilcote:2017fv} \\
\hline
$a$ (au) & & 10.2$^{+0.4}_{-0.3}$ & this work \\
$e$ & & 0.12$^{+0.04}_{-0.03}$ & this work \\
$i$ (deg) & & 88.88$\pm$0.09 & this work \\
$\omega$ (deg) & & 198$\pm 4$ & this work \\
$\Omega$ (deg) & & 32.05$\pm$0.07 & this work \\
$T_0$ & & 2013.7$\pm 0.2$ & this work \\
$P$ (yrs) & & 24.3$^{+1.5}_{-1.0}$ & this work \\
\enddata
\end{deluxetable*}

We refer to this orbit fit, to the \textit{Hipparcos} and \textit{Gaia} absolute astrometry, the CRIRES RV, and the relative astrometry from NACO, NICI, Magellan, and GPI, as ``Case 3.''  These results are presented in Figures~\ref{fig:triangle} and \ref{fig:tracks}, and Tables~\ref{tbl:system} and~\ref{tbl:posteriors}.  In this combined fit, the eccentricity has shifted upward slightly, with eccentricity $\lesssim0.05$ no longer allowed.  The other imaging parameters are similar to our previous imaging-only fit.  Astrometric parameters are similar to the \citet{vanLeeuwen:2007dc} \textit{Hipparcos} catalog values as well.  Offset from the \citet{vanLeeuwen:2007dc} reference location ($\alpha_{H0}^*$ and $\delta_{H0}$) is $0.06\pm0.11$ mas and $0.03\pm0.13$ mas, respectively.  Parallax of 51.44$\pm$0.13 mas is essentially the same as the \textit{Hipparcos} catalog value of 51.44$\pm$0.12 mas.  Meanwhile, as expected for significant reflex motion, we infer the proper motion of the system ($\mu_{\alpha^*}$, $\mu_\delta$) to be (+4.94$\pm 0.02$,+83.93$^{+0.03}_{-0.04}$) mas/yr, different from their catalog values of (+4.65$\pm$0.11, +83.10$\pm$0.15) mas/yr, by 2.2$\sigma$ and 4.4$\sigma$, respectively.

\begin{figure}[ht]
\includegraphics[width=\columnwidth,trim=2cm 2cm 2cm 6cm]{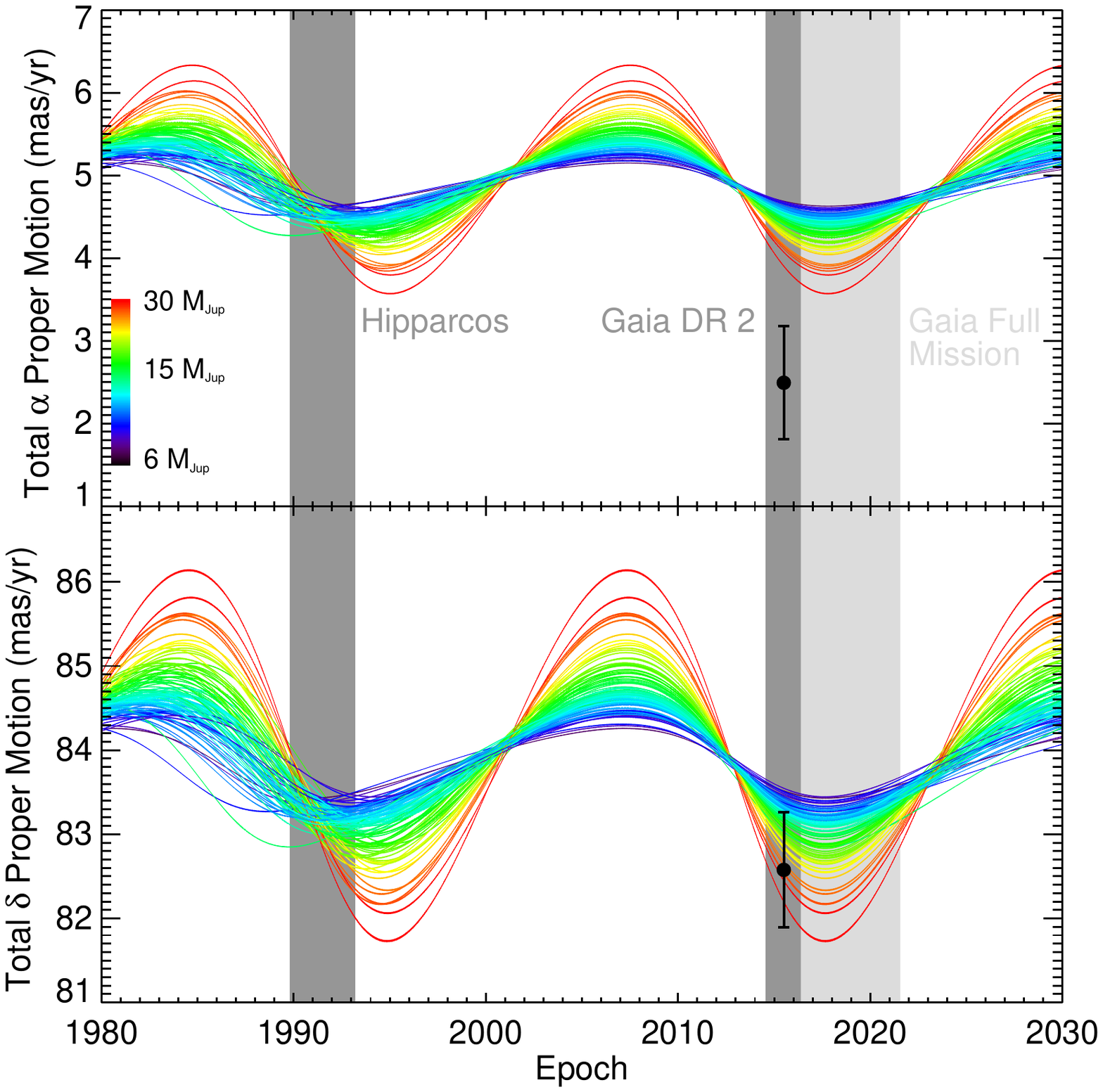}
\caption{The observed proper motion of $\beta$ Pic, including the system proper motion and the reflex motion due to the orbit of $\beta$ Pic b, with the tracks (color-coded by planet mass) drawn from the posterior, again using the orbit fit with all imaging data except SPHERE, as well as \textit{Hipparcos} and \textit{Gaia} (Case 3).  Dark gray bars mark the timeframe of the \textit{Hipparcos} and \textit{Gaia} observations, with the light gray bar representing the expected remaining extent of the full 7-year \textit{Gaia} mission.  The \textit{Hipparcos} IAD constrains the proper motion well between 1990-1993, and a more precise \textit{Gaia} proper motion measurement can greatly reduce the error bars on the planet mass. \label{fig:orbit_pm}}
\end{figure}

In Figure~\ref{fig:orbit_pm} we plot the predicted proper motion from our Case 3 orbit fit of $\beta$ Pic as a function of time.  The proper motion is well-constrained by the \textit{Hipparcos} IAD measurement between 1990--1993, and matches our accuracy on the system proper motion for this orbit fit ($+4.94 \pm 0.02$,$+83.93^{+0.03}_{-0.04}$) mas\,yr$^{-1}$.  Though we do not include the \textit{Gaia} DR2 proper motion measurement in this fit, we mark its location as points with error bars at 2015.5.  We note that the \textit{Gaia} DR2 proper motion errors ($\pm$0.68 mas/yr) are considerably larger than the \textit{Hipparcos} values of \citet{vanLeeuwen:2007dc}.  While the proper motion in declination is a good match to the tracks, the right ascension proper motion is significantly off from the tracks.  It is unclear if this is a result of systematics in extracting astrometry from bright stars, or whether this offset is the effect of attempting to fit an acceleration in proper motion over a 1.5 year time baseline with a 5-parameter fit.  If future \textit{Gaia} data releases are able to reach $<$0.1 mas/yr proper motion precision, it should greatly reduce the errors in the measurement of the mass of the planet.

\subsubsection{Independent analysis}
To probe the robustness of our results against different methods and algorithms we performed a second, independent analysis of the same dataset, in this case the dataset discussed above, as well as the SPHERE relative astrometry (referred to as "Case 5" below).  We reconstructed the HIP2 (\citealt{vanLeeuwen:2007book} IAD) abscissa using the method described in \citet[][Sect. 3.1]{Sahlmann:2011fk}. When fitting the standard linear 5-parameter model, we recovered the HIP2 catalog parameters and obtained a residual RMS of 0.79 mas. When adding the \textit{Gaia} DR2 position of \target\ (Gaia DR2 4792774797545105664) the RMS in the HIP2 residuals increases to 0.89 mas.  To correctly include the parallax-free \textit{Gaia} DR2 catalog position in the fit we set the corresponding parallax factors to zero.

In combination with the ground-based relative astrometry of \target\ b the \textit{Hipparcos} and \textit{Gaia} absolute astrometry allows us to determine model-independent dynamical masses of \target\ and its planetary companion, under the assumption that the space-based astrometry is unbiased (see next Section). We performed an MCMC analysis similar to \cite{Sahlmann:2016ab,Sahlmann:2013ab}. The 13 free parameters are $P$, $e$, $i$, $\omega$, $T_\mathrm{P}$, $M_\star$, $M_\mathrm{b}$,  $\Omega$ (8 parameters for the orbital motion) and $\alpha^\star_{2012}$, $\delta_{2012} $, $\varpi$, $\mu_{\alpha^\star}$, $\mu_\delta$ (5 parameters for the standard astrometric model), where we defined $\omega$ as the argument of periastron for the barycentric orbit of the primary (in the previous sections, $\omega$ referred to the relative orbit). In the MCMC we adjusted the pair $\sqrt{e}\cos{\omega}$ and $\sqrt{e}\sin{\omega}$ instead of $e$ and $\omega$ to mitigate the effect of correlations that naturally exists between those parameters. We also chose the reference epoch at year 2002 (between the \textit{Hipparcos} and \textit{Gaia} epochs) to mitigate correlations between positional offsets and proper motions.  Additionally, values of $\alpha_{2012}^*$ and $\delta_{2012}$ here correspond to the location of the $\beta$ Pic barycenter at the reference epoch, while in the previous fit $\alpha_{H0}^*$ and $\delta_{H0}$ referred to the location of the system photocenter at the reference epoch.  All priors are flat and seed values and their uncertainties for the MCMC chains were set based on either the 5-parameter fit above or previous orbital solutions. We used 160 walkers with 44000 steps each and discarded the first 25\% of samples, which yields more than 5 million samples per parameter.

The MCMC chains exhibit stable convergence and the posterior distributions show clearly peaked shapes. The residual RMS in the absolute astrometry (\textit{Hipparcos} and \textit{Gaia}) with the median orbital model is 0.80 mas, thus significantly smaller than the 0.89 mas obtained with the linear model. This confirms that orbital motion is detected in the absolute astrometry. 

\begin{figure}[ht]
\includegraphics[width=\columnwidth,trim=2cm 2cm 2cm 6cm]{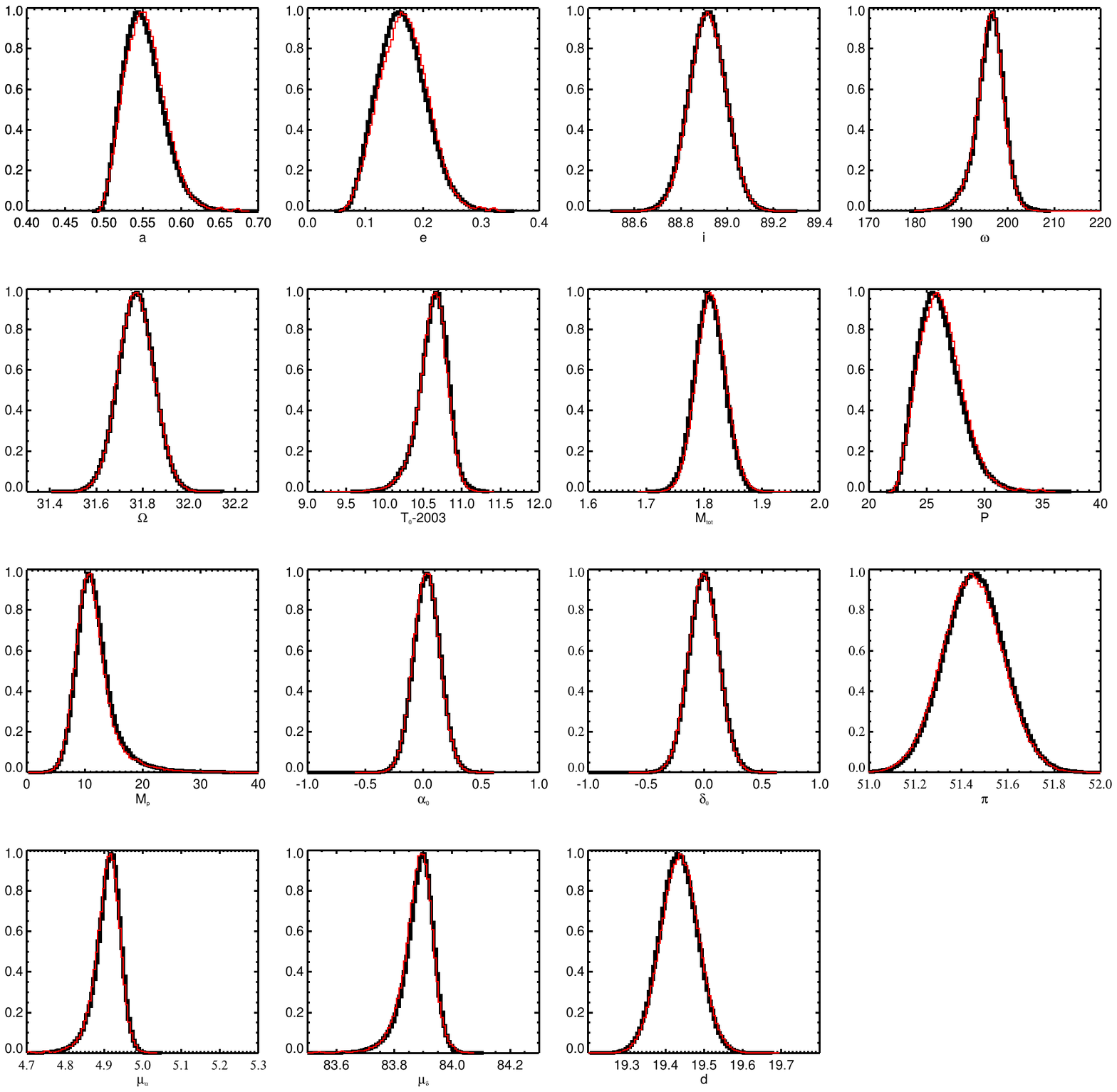}
\caption{Comparison of the two orbit-fitting techniques for Case 5 shows excellent agreement of the two sets of posteriors. \label{fig:compare_methods}}
\end{figure}

In terms of system parameters, the results of the two independent analyses are in excellent agreement as illustrated by Figure~\ref{fig:compare_methods}, with 1D posteriors overlapping.  This gives us confidence in the accuracy of both fitting algorithms.

\subsubsection{Bias in the Gaia DR2 catalog parameters due to orbital motion}
The source parameters in \textit{Gaia} DR2 were obtained by fitting either a 5-parameter model or a 2-parameter model to the astrometric data collected by the satellite \citep{Lindegren:2018aa}. For the 5-parameter solution of \target\ this means that any orbital motion present in the \textit{Gaia} astrometry was not accounted for specifically. Orbital motion may rather manifest itself as an increased excess noise or a bias in the DR2 parameters, which is worse for our purposes.

In an attempt to quantify the bias in the DR2 position caused by orbital motion, we simulated the individual \textit{Gaia} observations. We used the \textit{Gaia} Observation Forecast Tool (\emph{gost}, \url{https://gaia.esac.esa.int/gost/}) to predict the \textit{Gaia} focal plane crossings of \target\ in the timerange considered in DR2 after the ecliptic pole scanning (2014-08-22T21:00:00 -- 2016-05-23T11:35:00, \citealt{Lindegren:2018aa})\footnote{\cite{snellen:2018} mention a timerange between 2014-10-01 and 2016-04-19 for \target\ measurements.}. Unfortunately, the earliest date accepted by \emph{gost} is 2014-09-26T00:00:00, but we corrected for the missing month as described below. In the queried timerange, \emph{gost} predicted 26 \textit{Gaia} focal plane crossings in 16 visibility periods. The \textit{Gaia} DR2 catalog reports 30 \texttt{astrometric\_matched\_observations} in 15 visibility periods over the slightly longer timerange included in DR2. This validates that the \emph{gost} predictions are a reasonable approximation of the actual \textit{Gaia} observations. To account for the missing first month in the \emph{gost} prediction, we duplicated the last two \emph{gost} predictions and prepended them to the list of predictions with timestamps that correspond to the start of the DR2 timerange. Our simulated \textit{Gaia} observation setup this includes 28 focal plane crossings in 17 visibility periods. 

We used a set of the 13 parameters fitted in the previous section to compute noiseless \textit{Gaia} along-scan measurements (equivalent to the \textit{Hipparcos} abscissa) that include the orbital motion, setting the reference epoch to 2015.5. Observation times, parallax factors and scan angles were specified according to the \emph{gost} predictions. We also computed the model position of the star at epoch 2015 including barycentric orbital motion and proper motion (zero by definition of the reference epoch), but not parallax (by setting the parallax factor to zero) to replicate the parallax-free DR2 catalog position. 

We then fitted the standard 5-parameter linear model to the simulated \textit{Gaia} data of \target\ and compared the 2015.5 model position to the best-fit position offsets. The difference between the two corresponds to our estimate of the DR2 position bias. When no significant orbital motion is present (e.g. the planet mass is set to zero), both the model position at 2015.5 and the fitted coordinate offsets of the 5-parameter fit are zero and the input proper motions and parallax are recovered. When orbital motion is present, the actual and the linear-fit position are different. 

Since we cannot be certain about the fidelity of the \emph{gost} predicted DR2 epochs, we estimated the uncertainty in the bias estimation by repeating random draws of 28, 26, and 24 out of 28 predicted epochs. We also incorporated the varying fit parameters by using samples from the MCMC chains in the previous section. 

We considered three cases: Case (a): Nominal \textit{Gaia} DR2 positions and uncertainties, no bias correction;
Case (b): When drawing 28 or 26 epochs for the solution in the previous section and 10000 random draws and parameter sets, the DR2 coordinate bias due to orbital motion is estimated to $\epsilon_\mathrm{RA}=0.017\pm0.003$ mas and $\epsilon_\mathrm{Dec}=0.013\pm0.003$ mas, which is negligible given the DR2 position uncertainties of $\sim$0.3 mas; 
Case (c): If we draw 24 epochs, these estimates increase to $\epsilon_\mathrm{RA}=0.085\pm0.142$ mas and $\epsilon_\mathrm{Dec}=-0.040\pm0.142$ mas, so the bias essentially increases the uncertainty in the DR2 positions and introduces a minor shift. 

\begin{figure}[ht]
\includegraphics[width=\columnwidth]{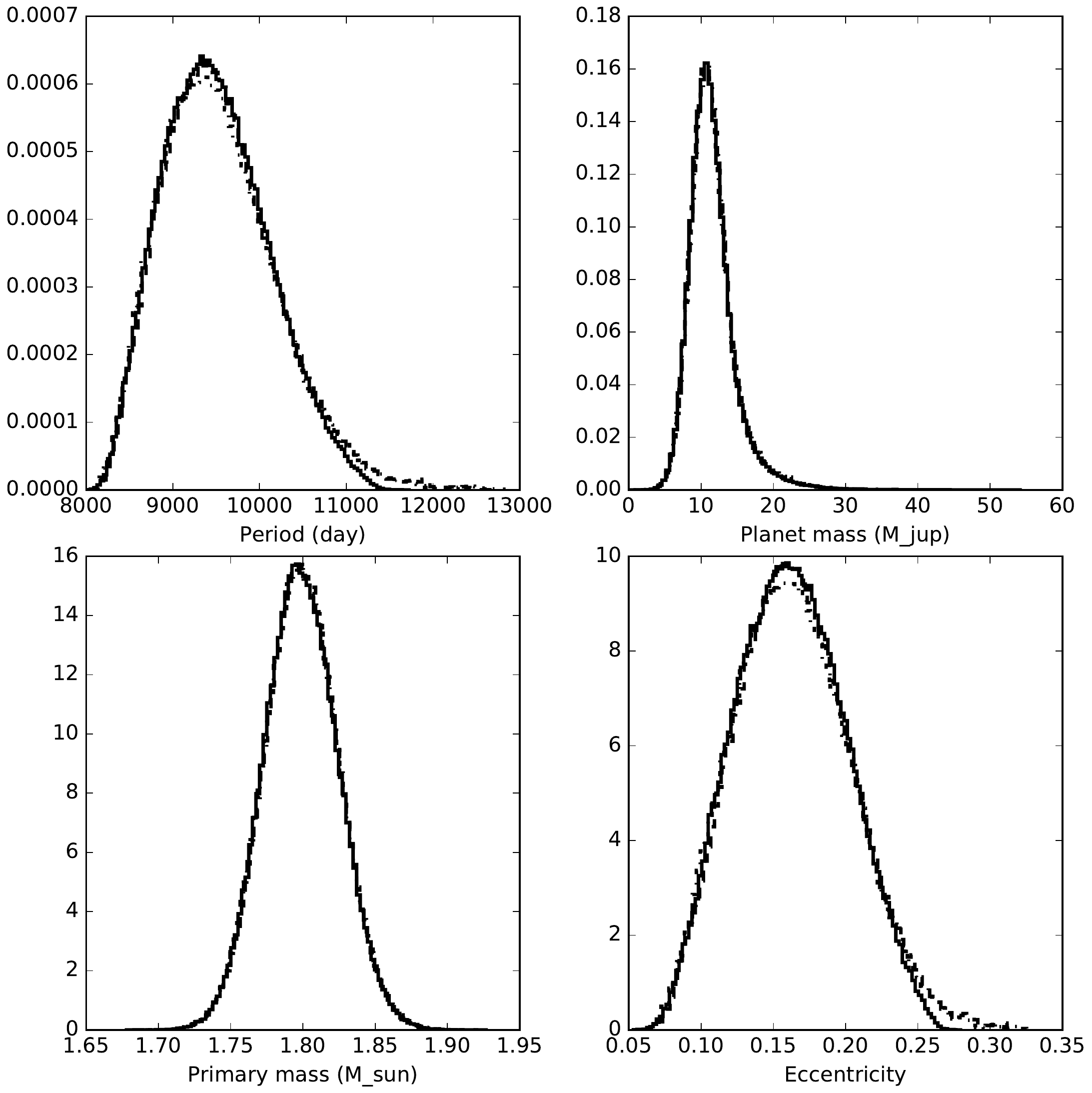}
\caption{Posteriors on star mass, planet mass, eccentricity, and period (yrs) with (dashed line: case (b), dash-dotted line: case (c)) and without (solid line: case (a)) DR2 position bias correction. \label{fig:effect_of_bias}}
\end{figure}

We repeated the MCMC analysis in all three cases. When debiasing the DR2 position, we subtracted $\epsilon$ from the catalog coordinate before including it in the fit and we added the bias uncertainty in quadrature to the DR2 position uncertainty. We found that the effect of the position bias as estimated above on the solution parameters is negligible. We illustrate this in Figure \ref{fig:effect_of_bias}, where we show posteriors for several fit parameters that are essentially indistinguishable. The same applies to all other parameters.

Whereas for our purposes the bias of the DR2 parameters due to orbital motion is negligible, this is certainly not the general case. For instance, we found that the bias in \target's DR2 proper motion is significant: $\epsilon_{\mu_{\alpha^\star}}=0.37\pm0.08$ mas/yr and $\epsilon_{\mu_\delta}=0.62\pm0.13$ mas/yr (the corresponding parallax bias is smaller than 3 $\mu$as).   A bias of $\sim$0.4 mas/yr in the RA direction is not enough to explain the offset seen in Figure~\ref{fig:orbit_pm}, where the \textit{Gaia} value of $\mu_{\alpha^\star}$ is $\sim$2 mas/yr from the orbit tracks, so the full cause of this offset is still unclear.  Likewise, the $\mu_{\delta}$ bias moves the \textit{Gaia} proper motion even further from the orbit tracks.   Caution is therefore necessary when using the DR2 parameters of systems exhibiting orbital motion, and in particular when determining orbital parameters from the Gaia DR2 catalog in combination with other surveys \citep[e.g.][]{Brandt:2018aa,Kervella:2018aa}.

\subsubsection{The effect of using different datasets}

We consider different combinations of relative astrometry to investigate how different combinations influence the derived mass.  In addition to the fit to \textit{Hipparcos}, \textit{Gaia}, CRIRES, NACO, NICI, Magellan, and GPI disussed above (``Case 3''), we also consider the effect of the 2018 GPI data by performing a second fit, but without these two GPI datapoints in 2018 (``Case 4'').  We perform three additional fits as well, all using the \textit{Hipparcos}, \textit{Gaia}, and CRIRES data: including the SPHERE data of \citet{lagrange:2018} (``Case 5'') as presented, including this SPHERE data but fitting for additional offset terms for the GPI separation and position angle (``Case 6''), and using relative astrometry only from ESO instruments, NACO and SPHERE (``Case 7'').  We present the full set of posteriors in Table~\ref{tbl:posteriors} for each of these orbit fits.

Given the small errors on the imaging data ($\sim$1 mas for the early GPI data), the importance of the astrometric calibration becomes key, especially when combining datasets from different instruments.  An uncorrected offset in calibration (either in plate scale or true north) results in a large acceleration between datapoints, which the orbit fitter will attempt to compensate for with a more eccentric orbit, where orbital speed can be varied near the problematic epochs.  It has been speculated that a calibration offset is the likely cause for different predictions for the Hill Sphere crossing and closest approach of $\beta$ Pic b in the previous two years \citep{Wang:2016gl}.  We examine the influence of this effect by performing multiple orbit fits combining imaging and the \textit{Hipparcos} and \textit{Gaia} data, with different combinations of instruments.

\begin{figure}[ht]
\includegraphics[width=\columnwidth,trim=2cm 3cm 2cm 17cm]{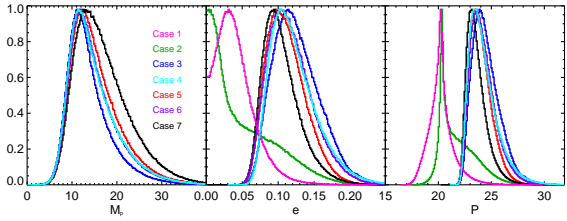}
\caption{Planet mass, eccentricity, and period posteriors for different datasets.  Datasets with GPI but not SPHERE data tend to favor smaller planet masses, and lager eccentricity and period (Cases 3 and 4).  Combinations of GPI and SPHERE data (5 and 6) have more probability at larger masses, and a fit that excluded GPI data (7) moves to the largest planet masses.\label{fig:massfig}}
\end{figure}

Figure~\ref{fig:massfig} shows posteriors for planet mass, eccentricity, and period for these multiple orbit fit cases.  Using GPI data but not SPHERE and also using $Hipparcos$ and $Gaia$ astrometry (Cases 3 and 4) give generally lower masses, with slightly larger eccentricity and period, compared to fits that incorporate SPHERE data.  The largest difference is from Case 3 (GPI but not SPHERE) with a mass of the planet of 12.8$^{+5.5}_{-3.2}$ M$_{\rm Jup}$ to Case 7 (SPHERE but not GPI), where the mass measurement is 15.8$^{+7.1}_{-4.7}$ M$_{\rm Jup}$, with combinations of the two instruments falling in between.

\begin{figure}[ht]
\includegraphics[width=\columnwidth,trim=2cm 3cm 2cm 7cm]{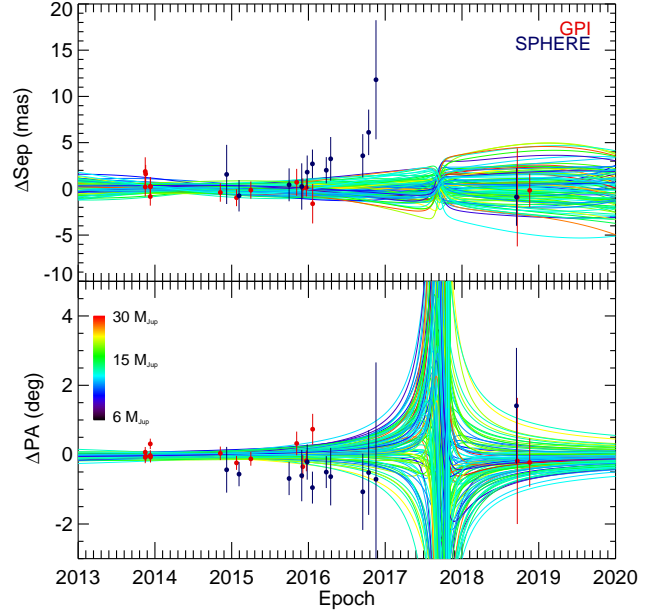}
\caption{Comparison of the GPI and SPHERE astrometry between 2013 and 2020, with the lowest $\chi^2$ orbit subtracted off (Case 3). We find no evidence for a systematic offset in plate scale ($1.001\pm0.003$), but significant evidence of a position angle offset of $-0.47 \pm 0.14^\circ$.  \label{fig:gpisphere}}
\end{figure}

Following the updates to the north angle in the GPI pipeline, we find evidence for a systematic position angle offset between GPI and SPHERE.  The Case 6 fit introduced two additional paraemters into the fit, a multiplicative offset to GPI separations, and an additive offset to GPI position angles (corresponding to calibration errors in planet scale and true north, respectively).  The fit values for these offsets are $\rho_S / \rho_G$ = 1.001 $\pm$ 0.003, and $\theta_S - \theta_G$ = -0.47 $\pm$ 0.14$^\circ$, suggesting no offset in plate scale, but a true north offset of about half a degree between the two instruments.  Figure~\ref{fig:gpisphere} compares the relative astrometry from GPI and SPHERE, indeed showing SPHERE position angles systematically $\sim$0.5$^\circ$ smaller than GPI data at the same epoch.

\citet{maire:2019} present new astrometry of the planet 51 Eri b from SPHERE and an independent reduction of GPI data, and find from their analysis a systematic PA offset of 1.0$\pm$0.2$^\circ$, with SPHERE having larger values than GPI.  With the revised astromeric calibration, we found an offset of $\Delta\theta = -0.16\pm0.26$\,deg from a joint fit to our reduction of our GPI data and the SPHERE astrometry published in \citet{maire:2019} (De Rosa et al. 2019, \textit{submitted}), consistent with the offset found for $\beta$~Pic~b in this work. Using the old astrometric calibration and data reduced with the same version of the DRP used by \citet{maire:2019} we calculated an offset of $\Delta\theta=0.28\pm0.26$\,deg, closer to the value in \citet{maire:2019}, but still significantly different.  This lends further evidence to the conclusion that the culprit is not a single constant offset between the two instruments, but perhaps an algorithmic difference in how astrometry is extracted. Indeed, \citet{maire:2019} note that when they refit GPI data on 51 Eri, they find $\sim0.35^\circ$ larger values of PA than those presented by \citet{DeRosa:2015jl} for the same datasets. Further analysis is ongoing to determine the precise cause of these offsets, and their impact on derived orbital parameters.

\subsubsection{Comparison to previous orbit fits}

\begin{figure}[ht]
\includegraphics[width=\columnwidth,trim=2cm 3cm 2cm 8cm]{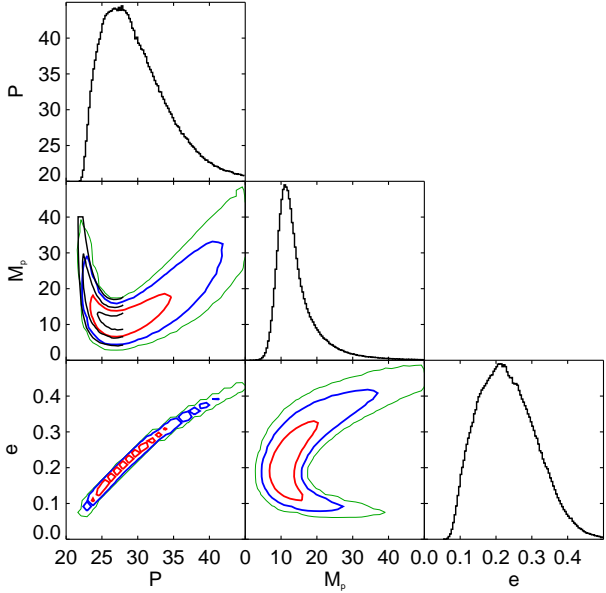}
\caption{Posteriors from the combined imaging and astrometric fit for period, planet mass, and eccentricity, but without the 2018 GPI data (Case 4).  Overplotted in the mass/period covariance plot are 1, 2, and 3$\sigma$ contours extracted from Fig 3 of \citet{snellen:2018} for the same dataset.  While we find generally good agreement with covariance contours for periods less than 28 years, there is significant probability at larger periods and masses, creating a more uncertain mass measurement ($12.7^{+6.4}_{-3.1} $ M$_{\rm Jup}$) than reported by \citet{snellen:2018} (11 $\pm$ 2 M$_{\rm Jup}$). \label{fig:small_triangle}}
\end{figure}

In Figure~\ref{fig:small_triangle}, we compare a modified vesion of our Case 4 to the results from \citet{snellen:2018}, who examined a similar relative astrometric dataset.  For consistency in this comparison, here we use the published astrometry from \citet{Wang:2016gl}, rather than the updated astrometry presented here.  We also do not use the CRIRES RV or the 2009 NaCo $M$-band point for this fit, to match the \citet{Wang:2016gl} orbit fitting.  Key differences in the method is that \citet{snellen:2018} did not simultaneously fit the relative astrometry and \textit{Hipparcos} and \textit{Gaia} data as we did, but rather took the \citet{Wang:2016gl} orbital element posteriors as the constraints from the relative astrometric fit.  Additionally, while we use the \textit{Hipparcos} IAD as constraints on the orbit in the plane of the sky, \citet{snellen:2018} converted the IAD into one-dimensional measurements along the orbital plane given by \citet{Wang:2016gl}.  When reporting the mass posterior, \citet{snellen:2018} restricted the fit to the 1$\sigma$ period range of \citet{Wang:2016gl}, the most circular orbits.  But as seen in Table~\ref{tbl:posteriors}, the addition of the \textit{Hipparcos} and \textit{Gaia} data push the visual orbit toward longer periods and higher eccentricity than the \citet{Wang:2016gl} fit to the relative astrometry alone.

As a result, while \citet{snellen:2018} find a well constrained mass for the planet of 11 $\pm$ 2 M$_{\rm Jup}$, we find a broader range of 12.7$^{+6.4}_{-3.1}$ M$_{Jup}$ when analyzing the same dataset.  A key factor in the reported smaller uncertainty is that \citet{snellen:2018} fixed a number of parameters, including the mass of the star ($M_{tot}$) and position angle of nodes ($\Omega$), as well as the orbital period of the planet.  In the covariance panel between mass and period within the triangle plot of Figure~\ref{fig:small_triangle}, we show the \citet{snellen:2018} 1, 2, and 3 $\sigma$ contours, extracted from their Figure 3, against ours.  By restricting the period range to the 1 $\sigma$ range of \citet{Wang:2016gl} of $<$28 years, the planet mass appears more constrained than it actually is given the full dataset.  Including the GPI 2018 astrometry, as well as updating the astrometry following fixes to the pipeline, (Case 3) produces a somewhat more constrained planet mass compared to our modified Case 4, 12.8$^{+5.5}_{-3.2}$ M$_{\rm Jup}$, but with error bars still a factor of two larger than reported by \citet{snellen:2018}.  This offset illustrates the importance of a simultaneous fit of relative and absolute astrometry, given the complicated covariant structure of such orbits.

Recently, \citet{dupuy:2019} presented a fit to the $\beta$ Pic b orbit based on relative astrometry from the literature (including the \citet{lagrange:2018} SPHERE measurement from 2018) and the \textit{Hipparcos}-\textit{Gaia} Catalog of Accelerations (HGCA, \citealt{Brandt:2018aa}).  Their analysis differs from ours in a number of ways; most significantly, they utilize the \textit{Hipparcos} catalog values rather than the \textit{Hipparcos} IAD and their fit includes the \textit{Gaia} proper motion for $\beta$ Pic, though with inflated errors.  Additionally, 
our analysis benefits from the more precise relative astrometry from GPI in 2018.  \citet{dupuy:2019} also fit the radial velocity (RV) of the star \citep{Lagrange:2012} and of the planet \citep{snellen:2014}, though given the large jitter in the stellar RVs and the moderate error bars on the planet RV, we don't expect the inclusion of RVs to have a significant difference in the two fits.  We also find a more constrained parallax for the system (51.44 $\pm$ 0.13 mas from our Case 3 fit, largely based on the \textit{Hipparcos} IAD), compared to their inflated \textit{Hipparcos} parallax error, a linear combination of the original \citet{ESA:1997ws} catalog and the re-reduced \citet{vanLeeuwen:2007dc} catalog.

Similar to the comparison to \citet{snellen:2018}, we produce a modified Case 3 fit, before the correction to the GPI astrometry, to compare the two methods. \citet{dupuy:2019} find orbital parameters generally similar to our modified Case 3 orbit fit.  They find a planet mass of $13.1^{+2.8}_{-3.2}$ M$_{\rm Jup}$, period of $29.9^{+2.9}_{-3.2}$ yrs, and eccentricity of 0.24 $\pm$ 0.06, compared to our values of $11.1^{+2.7}_{-2.3}$ M$_{\rm Jup}$, 27.1 $\pm$ 2.0 yrs, and 0.19 $\pm$ 0.05.  Thus we find a somewhat lower planet mass, shorter period, and smaller eccentricity, with slightly smaller error bars.  We find a stellar mass of 1.81 $\pm$ 0.03 M$_\odot$, similar to the \citet{dupuy:2019} value of 1.84 $\pm$ 0.05 M$_\odot$; these two estimates are the first time planet mass and stellar mass have been measured simultaneously from the same fit for a directly imaged planet.  Comparing our triangle plot in Figure~\ref{fig:triangle} to their Figure 3, \citet{dupuy:2019} do not reproduce our U-shaped covariance between
semi-major axis and planet mass, and eccentricity and planet mass, rather they see a roughly linear relationship for both covariances.  The intersection of the two sets of covariances includes short period lower-mass planets and long period higher-mass planets, while our results include another family of short period higher-mass planets not seen by \citet{dupuy:2019}.  The source of this discrepency is not clear: given the large error bars on both the RVs and the recomputed Gaia proper motion of \citet{dupuy:2019}, neither should significantly move the fit.  It is not likely that the GPI 2018 data is to blame, since our Case 7 fit (which is a similar imaging dataset to the one used by \citealt{dupuy:2019}) also has these U-shaped covariances.  For our final Case 3 fit, using updated GPI astrometry, we mass of  12.8$^{+5.5}_{-3.2}$ M$_{\rm Jup}$, eccentricity of 0.12$^{+0.04}_{-0.03}$, and period of 24.3$^{+1.5}_{-1.0}$ yrs: larger uncertainty on planet mass, and smaller values of period and eccentricity.

\begin{longrotatetable}
\begin{deluxetable*}{ccccccccccccccccc}
\tablewidth{0pt}
\tabletypesize{\tiny}
\tablecaption{Orbit Fitting Posteriors\label{tbl:posteriors}}
\tablehead{ \colhead{  } & \colhead{a (au)} & \colhead{e} & \colhead{i (deg)} & \colhead{$\omega$ (deg)} & \colhead{$\Omega$ (deg)} & \colhead{T$_0$} & \colhead{P (yr)} & \colhead{M$_{\textrm{tot}}$} & \colhead{M$_{\textrm{P}}$ (M$_{\textrm{Jup}}$)} & \colhead{$\alpha_{H0}^*$ (mas)} & \colhead{$\delta_{H0}^*$ (mas)} & \colhead{$\pi$ (mas)} & \colhead{$\mu_{\alpha^*}$} & \colhead{$\mu_\delta$} & \colhead{$\rho_{S} / \rho_{G}$} & \colhead{$\theta_S - \theta_G$ (deg)}}
\startdata
\hline
\multicolumn{17}{c}{Case 1: NACO, NICI, MagAO, GPI, prior to 2018}\\
\hline
Median & 8.95 & 0.036 & 88.803 & 297.33 & 32.025 & 2018.55 & 20.2 & 1.75 & & & & & & & & \\
68\% CL min. & 8.63 & 0.014 & 88.684 & 222.43 & 31.939 & 2014.59 & 19.2 & 1.72 & & & & & & & & \\
68\% CL max. & 9.25 & 0.065 & 88.921 & 355.43 & 32.110 & 2021.47 & 21.2 & 1.78 & & & & & & & & \\
95\% CL min. & 8.30 & 0.002 & 88.564 & 194.72 & 31.853 & 2009.06 & 18.2 & 1.69 & & & & & & & & \\
95\% CL max. & 9.87 & 0.106 & 89.038 & 478.25 & 32.193 & 2023.33 & 23.3 & 1.81 & & & & & & & & \\
\hline
\multicolumn{17}{c}{Case 2: NACO, NICI, MagAO, GPI}\\
\hline
Median & 9.32 & 0.038 & 88.835 & 203.63 & 32.005 & 2013.49 & 21.3 & 1.78 & & & & & & & & \\
68\% CL min. & 9.01 & 0.009 & 88.734 & 192.91 & 31.924 & 2010.05 & 20.3 & 1.76 & & & & & & & & \\
68\% CL max. & 9.96 & 0.101 & 88.935 & 475.83 & 32.084 & 2014.23 & 23.5 & 1.81 & & & & & & & & \\
95\% CL min. & 8.81 & 0.001 & 88.633 & 183.19 & 31.843 & 2005.52 & 19.5 & 1.74 & & & & & & & & \\
95\% CL max. & 10.59 & 0.158 & 89.034 & 535.92 & 32.163 & 2024.12 & 25.8 & 1.83 & & & & & & & & \\
\hline
\multicolumn{17}{c}{\textbf{Case 3: NACO, NICI, MagAO, GPI, Hipparcos, Gaia (Adopted Parameters)}}\\
\hline
Median & \textbf{10.18} & \textbf{0.122} & \textbf{88.877} & \textbf{18.12} & \textbf{-147.954} & \textbf{2013.72} & \textbf{24.3} & \textbf{1.79} & \textbf{12.82} & \textbf{0.057} & \textbf{0.034} & \textbf{51.439} & \textbf{4.936} & \textbf{83.927}& & \\
68\% CL min. & \textbf{9.89} & \textbf{0.094} & \textbf{88.782} & \textbf{14.09} & \textbf{-148.029} & \textbf{2013.47} & \textbf{23.3} & \textbf{1.76} & \textbf{9.58} & \textbf{-0.058} & \textbf{-0.096} & \textbf{51.309} & \textbf{4.911} & \textbf{83.890}& & \\
68\% CL max. & \textbf{10.60} & \textbf{0.159} & \textbf{88.971} & \textbf{22.04} & \textbf{-147.880} & \textbf{2013.95} & \textbf{25.8} & \textbf{1.81} & \textbf{18.29} & \textbf{0.173} & \textbf{0.166} & \textbf{51.570} & \textbf{4.960} & \textbf{83.962}& & \\
95\% CL min. & \textbf{9.69} & \textbf{0.074} & \textbf{88.687} & \textbf{9.16} & \textbf{-148.103} & \textbf{2013.17} & \textbf{22.6} & \textbf{1.73} & \textbf{7.07} & \textbf{-0.173} & \textbf{-0.226} & \textbf{51.177} & \textbf{4.880} & \textbf{83.841}& & \\
95\% CL max. & \textbf{11.10} & \textbf{0.199} & \textbf{89.064} & \textbf{26.32} & \textbf{-147.806} & \textbf{2014.19} & \textbf{27.6} & \textbf{1.84} & \textbf{27.07} & \textbf{0.290} & \textbf{0.298} & \textbf{51.700} & \textbf{4.984} & \textbf{84.000}& & \\
\hline
\multicolumn{17}{c}{Case 4: NACO, NICI, MagAO, GPI, Hipparcos, Gaia, prior to 2018}\\
\hline
Median & 10.07 & 0.116 & 88.871 & 22.28 & -147.943 & 2013.96 & 24.0 & 1.77 & 13.61 & 0.065 & 0.045 & 51.437 & 4.941 & 83.936& & \\
68\% CL min. & 9.78 & 0.091 & 88.760 & 14.72 & -148.026 & 2013.52 & 23.0 & 1.74 & 9.91 & -0.051 & -0.087 & 51.306 & 4.915 & 83.896& & \\
68\% CL max. & 10.52 & 0.154 & 88.981 & 31.35 & -147.862 & 2014.48 & 25.6 & 1.81 & 20.02 & 0.182 & 0.178 & 51.567 & 4.965 & 83.972& & \\
95\% CL min. & 9.60 & 0.074 & 88.649 & 8.05 & -148.108 & 2013.10 & 22.5 & 1.71 & 7.24 & -0.167 & -0.218 & 51.176 & 4.878 & 83.837& & \\
95\% CL max. & 11.17 & 0.202 & 89.090 & 41.66 & -147.781 & 2015.06 & 27.8 & 1.84 & 29.28 & 0.300 & 0.312 & 51.699 & 4.989 & 84.010& & \\
\hline
\multicolumn{17}{c}{Case 5: NACO, NICI, MagAO, GPI, SPHERE, Hipparcos, Gaia}\\
\hline
Median & 10.03 & 0.108 & 89.003 & 17.81 & -147.921 & 2013.69 & 23.9 & 1.77 & 14.25 & 0.069 & 0.050 & 51.436 & 4.944 & 83.940& & \\
68\% CL min. & 9.79 & 0.085 & 88.925 & 13.23 & -147.991 & 2013.41 & 23.0 & 1.75 & 10.32 & -0.047 & -0.081 & 51.305 & 4.921 & 83.907& & \\
68\% CL max. & 10.36 & 0.139 & 89.081 & 21.94 & -147.850 & 2013.94 & 25.0 & 1.80 & 20.60 & 0.185 & 0.181 & 51.567 & 4.966 & 83.974& & \\
95\% CL min. & 9.63 & 0.069 & 88.846 & 8.12 & -148.062 & 2013.10 & 22.5 & 1.72 & 7.51 & -0.162 & -0.211 & 51.174 & 4.896 & 83.869& & \\
95\% CL max. & 10.77 & 0.174 & 89.159 & 27.09 & -147.780 & 2014.23 & 26.5 & 1.83 & 29.69 & 0.302 & 0.314 & 51.698 & 4.990 & 84.011& & \\
\hline
\multicolumn{17}{c}{Case 6: NACO, NICI, MagAO, GPI, SPHERE, Hipparcos, Gaia, with calibration offsets}\\
\hline
Median & 10.10 & 0.114 & 88.986 & 17.82 & -148.337 & 2013.70 & 24.0 & 1.78 & 13.71 & 0.063 & 0.044 & 51.436 & 4.941 & 83.935 & 1.001 & -0.471\\
68\% CL min. & 9.82 & 0.087 & 88.906 & 13.78 & -148.483 & 2013.45 & 23.1 & 1.75 & 10.04 & -0.053 & -0.086 & 51.304 & 4.917 & 83.899 & 0.998 & -0.613\\
68\% CL max. & 10.49 & 0.148 & 89.064 & 21.84 & -148.192 & 2013.94 & 25.4 & 1.82 & 19.90 & 0.179 & 0.176 & 51.568 & 4.964 & 83.970 & 1.004 & -0.329\\
95\% CL min. & 9.62 & 0.069 & 88.827 & 8.84 & -148.629 & 2013.15 & 22.5 & 1.72 & 7.34 & -0.168 & -0.217 & 51.173 & 4.888 & 83.852 & 0.996 & -0.757\\
95\% CL max. & 11.01 & 0.190 & 89.143 & 26.77 & -148.046 & 2014.21 & 27.2 & 1.85 & 29.06 & 0.297 & 0.309 & 51.699 & 4.988 & 84.009 & 1.007 & -0.186\\
\hline
\multicolumn{17}{c}{Case 7: NACO, SPHERE, Hipparcos, Gaia}\\
\hline
Median & 9.82 & 0.101 & 89.115 & 23.34 & -148.111 & 2014.03 & 23.5 & 1.72 & 15.77 & 0.082 & 0.068 & 51.431 & 4.951 & 83.952 & & \\
68\% CL min. & 9.57 & 0.081 & 88.953 & 13.63 & -148.376 & 2013.44 & 22.8 & 1.65 & 11.06 & -0.035 & -0.064 & 51.300 & 4.928 & 83.918 & & \\
68\% CL max. & 10.18 & 0.130 & 89.277 & 34.22 & -147.847 & 2014.65 & 24.6 & 1.79 & 22.89 & 0.199 & 0.201 & 51.561 & 4.973 & 83.985 & & \\
95\% CL min. & 9.38 & 0.066 & 88.788 & 4.96 & -148.641 & 2012.82 & 22.3 & 1.59 & 7.86 & -0.151 & -0.196 & 51.169 & 4.902 & 83.877 & & \\
95\% CL max. & 10.71 & 0.170 & 89.436 & 48.57 & -147.583 & 2015.31 & 26.2 & 1.86 & 32.21 & 0.316 & 0.335 & 51.692 & 4.996 & 84.022 & & \\
\enddata
\end{deluxetable*}
\end{longrotatetable}

\subsection{Effects of a second giant planet in the $\beta$ Pic system}

After this paper was submitted, \citet{lagrange:2019} presented radial velocity measurements of $\beta$ Pic and an orbit fit for an inner giant planet, $\beta$ Pic c, orbiting at 2.7 au.  In particular, by fitting for the $\delta$ Scuti pulsations of the star, they were able to detect the $\sim$4 year signal of the inner planet.  The orbit fit performed took 219 sets of orbital elements from a chain of a separate MCMC orbit fit to the astrometry, and used these elements as the basis for fitting the RVs of the star, with the added assumption that the two planets are coplanar.

Here we perform a joint fit to four types of data simulaneously: imaging data from NaCo, NICI, MagAO, and GPI, absolute astrometry from $Hipparcos$ and $Gaia$, radial velocity of the planet from CRIRES (the three of which constitute Case 3 above), as well as the $\delta$ Scuti-corrected RVs and errors of the star from \citet{lagrange:2019} (their supplementary Table 1).  We expand our 13-element orbit fit with an additional 8 parameters: semi-major axis, eccentricity, inclination angle, argument of periastron, position angle of nodes, epoch of periastron passage, and planet mass of $\beta$ Pic c, along with the RV offset of the star.  As with $\beta$ Pic b, we place priors on $\beta$ Pic c orbital parameters that are uniform in: log($a$), in eccentricity ($e$), in cosine of the inclination angle ($\cos{i}$), in argument of periastron ($\omega$), in position angle of nodes ($\Omega$), in epoch of periastron passage ($T_0$), in planet mass ($M_{c}$), and in RV offset ($\gamma$).  In addition to these priors, we perform a second "coplanar" fit, where the mutual inclination ($i_m$) between the two planets given by:

\begin{equation}
    \cos{i_\textrm{m}} = \cos{i_\textrm{b}} \cos{i_\textrm{c}} + \sin{i_\textrm{b}} \cos{i_\textrm{c}} \cos{(\Omega_\textrm{b} - \Omega_\textrm{c})}
\end{equation}

\noindent is constrained to be a Gaussian centered on 0 with standard deviation of 1$^\circ$.  In both cases, the two planets are assumed to not interact with each other, so that position and radial velocity of the star is just the linear combination of the reflex motion from each planet's orbit.  We also note that the RV offset $\gamma$ does not represent the system velocity, since  \citet{lagrange:2019} have subtracted off the $\delta$ Scuti pulsations, and so any RV offset.  Here, $\gamma$ represents an additional RV correction beyond this. 

\begin{figure*}[ht]
\includegraphics[width=\textwidth,trim=2cm 4cm 2cm 9cm]{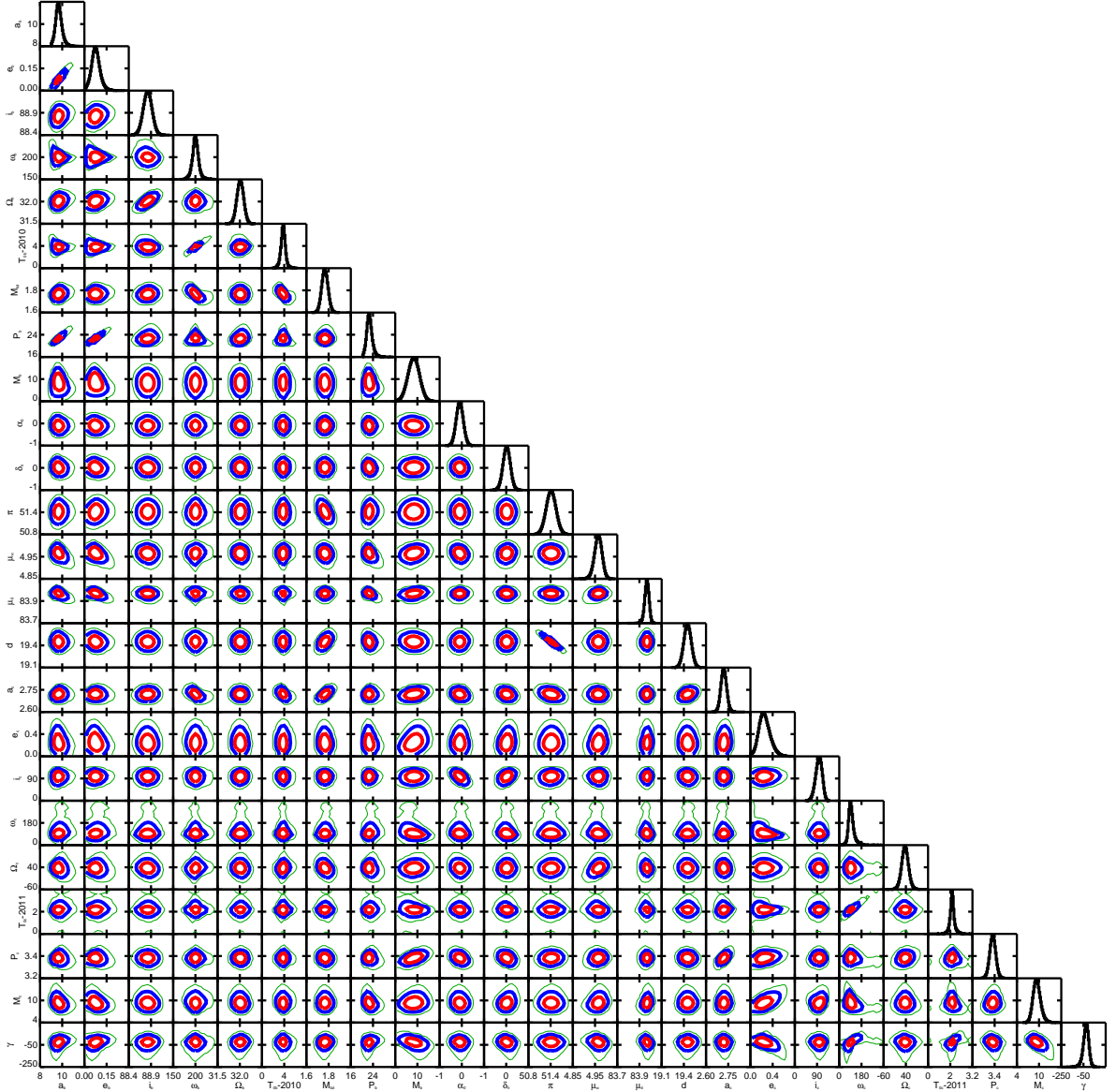}
\caption{Posteriors from the fit to Case 3 and the stellar RVs from \citet{lagrange:2019}, including a second planet ($\beta$ Pic c), with no additional constraints on mutual inclination between the two planets. \label{fig:ctriangle}}
\end{figure*}

We give the posteriors to the unconstrained mutual inclination fit in Figure~\ref{fig:ctriangle} and in Table~\ref{tbl:posteriors2}. Despite having no constraint on mutual inclination, the inclination angle and position angle of nodes for c ($i_\textrm{c}$ and $\Omega_\textrm{c}$) differ from the priors, and follow the orbit of $\beta$ Pic b, but with larger uncertainties: $i_\textrm{b} = 88.8 \pm 1.0^\circ$ and $\Omega_\textrm{b} = 32.02 \pm 0.08^\circ$ for the outer planet, compared to $i_\textrm{c} = {98^{+12}_{-14}}^\circ$ and $\Omega_\textrm{c} = 36 \pm 15^\circ$ for the inner planet.  Since radial velocities do not constrain either of these parameters, the absolute astrometry of $Hipparcos$ and $Gaia$ must be supplying these constraints.

\begin{figure*}[ht]
\includegraphics[width=\textwidth,trim=2cm 3.5cm 2cm 21cm]{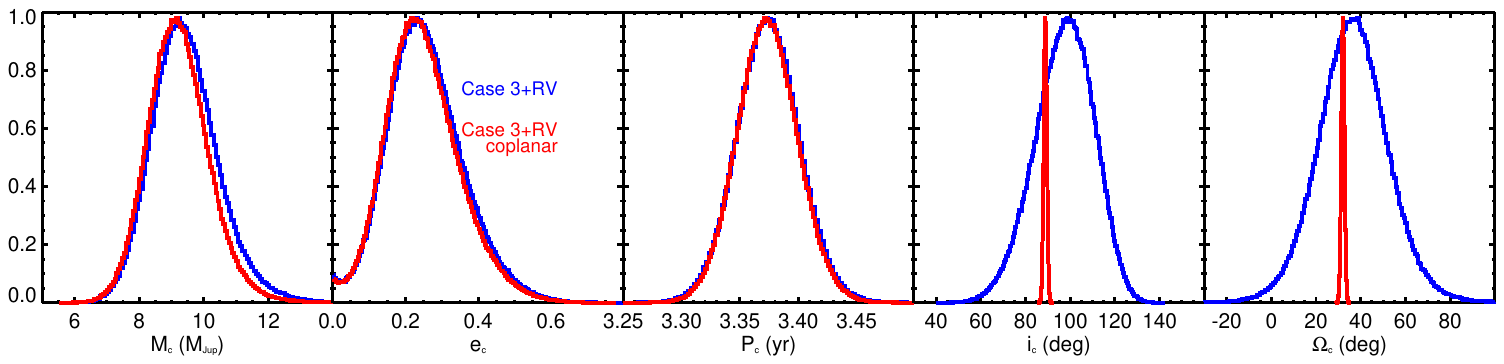}
\caption{Comparison of posteriors on the orbital parameters on the inner planet, $\beta$ Pic c, with unconstrained mutual inclination angle (blue) and a coplanar fit (red).  While the two fits differ greatly in the derived inclination angle and position angle of nodes, the other parameters are very similar.  The coplanar fit favors slightly smaller planet masses, M$_\textrm{c} = 9.4^{+1.1}_{-0.9}$  M$_\textrm{Jup}$ for the unconstrained mutual inclination fit, and M$_\textrm{c} = 9.2^{+1.0}_{-0.9}$  M$_\textrm{Jup}$ for the coplanar fit. \label{fig:coplanar}}
\end{figure*}

\begin{figure*}[ht]
\includegraphics[width=\textwidth,trim=2cm 3.5cm 2cm 21cm]{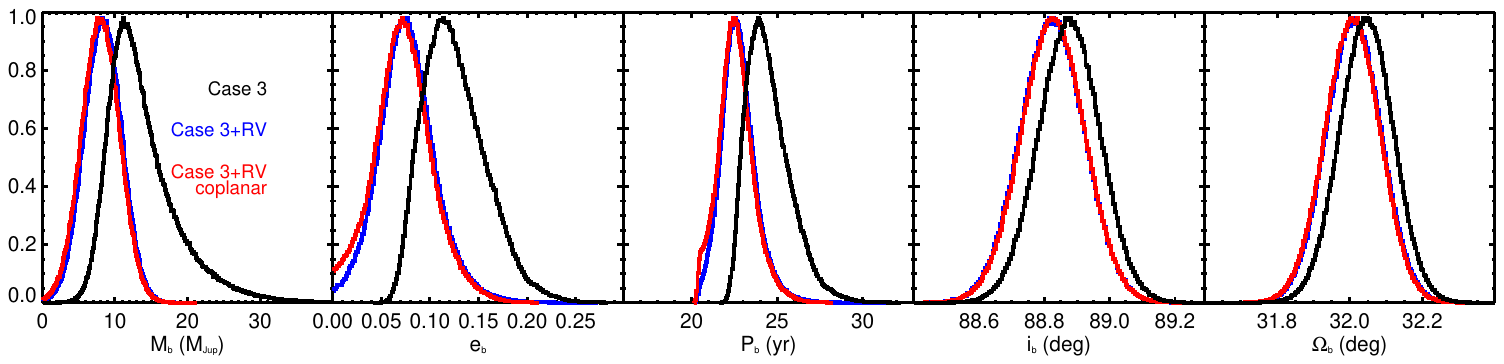}
\caption{The parameters of the outer planet $\beta$ Pic b, for the unconstrained mutual inclination fit (blue), the coplanar fit (red), and the regular Case 3 fit assuming only one planet in the system (black).  The addition of the radial velocities and a second planet push the mass of $\beta$ Pic b to lower values, as well as slightly decreasing eccentricity, period, inclination angle, and position angle of nodes. \label{fig:coplanar2}}
\end{figure*}

Other than inclination angle and position angle of nodes for the inner planet, there are no significant differences in the derived posteriors for the parameters of $\beta$ Pic c between the two fits.  The mass of $\beta$ Pic c changes slightly: in the fit that does not constrain mutual inclination it is M$_\textrm{c} = 9.4^{+1.1}_{-0.9}$  M$_\textrm{Jup}$, compared to M$_\textrm{c} = 9.2^{+1.0}_{-0.9}$  M$_\textrm{Jup}$ in the coplanar fit (Figure~\ref{fig:coplanar}).  This is true also for the parameters of the outer planet, $\beta$ Pic b, as shown in Figure~\ref{fig:coplanar2}, where the two fits incorporating two planets have similar posteriors on $\beta$ Pic b.

Similarly to \citet{lagrange:2019}, we find the presence of the c planet results in a lower mass for the b planet.  In our one-planet Case 3 fit, we found a mass for $\beta$ Pic b of M$_\textrm{b} = 12.8^{+5.5}_{-3.2}$  M$_\textrm{Jup}$, which drops to M$_\textrm{b} = 8.0 \pm 2.6$ M$_\textrm{Jup}$ in the coplanar fit (Figure~\ref{fig:coplanar2}).  In this coplanar fit the semi-major axis, period, and eccentricity posteriors also shift to lower values compared to the one-planet Case 3 fit.  A possible explanation for this is that the evidence for a non-zero eccentricity of $\beta$ Pic b came from the absolute astrometry of the star, and that this astrometric motion can be equally well explained with a more circular outer planet and a second inner planet.

Compared to \citet{lagrange:2019}, we find generally similar values to our fit, but with noticeable differences, likely resulting from performing a joint fit for all data and both planets, rather than using MCMC chains from a fit to $\beta$ Pic b to fit the RVs.  \citet{lagrange:2019} found values of [$a_\textrm{c}$, $e_\textrm{c}$, P$_\textrm{c}$, M$_\textrm{c}$] of [$2.69 \pm 0.003$ au, $0.24 \pm 0.02$, $3.335 \pm 0.005$ yr, $8.93 \pm 0.14$  M$_\textrm{Jup}$]
compared to values from our coplanar fit of [$2.72 \pm 0.019$ au, $0.24^{+0.1}_{-0.09}$, $3.39 \pm 0.02$ yr, $9.18^{+1.0}_{-0.9}$ M$_\textrm{Jup}$].  That these measurements are in such good agreement, but with errors several times larger from the joint fit, suggests that extracting a limited number of orbits from the posterior, as done in \citet{lagrange:2019}, underestimates the errors on the derived parameters.

\begin{deluxetable*}{l|ccccc|ccccc}
\tablewidth{0pt}
\tabletypesize{\small}
\tablecaption{Two-planet Fit Posteriors\label{tbl:posteriors2}}
\tablehead{  \colhead{ } & \multicolumn{5}{c}{Unconstrained mutual inclination} & \multicolumn{5}{c}{Coplanar} \\
\colhead{} & \colhead{} & \multicolumn{2}{c}{68\% CL} & \multicolumn{2}{c}{95\% CL} & \colhead{} & \multicolumn{2}{c}{68\% CL} & \multicolumn{2}{c}{95\% CL} \\
\colhead{  } & \colhead{Median} & \colhead{min.} & \colhead{max.} & \colhead{min.} & \colhead{max.} & \colhead{Median} & \colhead{min.} & \colhead{max.} & \colhead{min.} & \colhead{max.} }
\startdata
a$_{\textrm{b}}$ (au) & 9.68 & 9.43 & 9.98 & 9.16 & 10.40 & 9.65 & 9.39 & 9.95 & 9.10 & 10.34\\
e$_{\textrm{b}}$ & 0.076 & 0.049 & 0.105 & 0.020 & 0.143 & 0.072 & 0.043 & 0.101 & 0.012 & 0.138\\
i$_{\textrm{b}}$ (deg) & 88.824 & 88.726 & 88.922 & 88.627 & 89.019 & 88.826 & 88.729 & 88.923 & 88.630 & 89.019\\
$\omega_{\textrm{b}}$ (deg) & -159.79 & -166.32 & -152.88 & -174.29 & -139.61 & -160.13 & -167.09 & -152.40 & -175.46 & 166.13\\
$\Omega_{\textrm{b}}$ (deg) & 32.011 & 31.934 & 32.087 & 31.858 & 32.162 & 32.008 & 31.932 & 32.084 & 31.855 & 32.159\\
T$_\textrm{0b}$ & 2013.81 & 2013.41 & 2014.20 & 2012.89 & 2014.81 & 2013.77 & 2013.32 & 2014.17 & 2012.47 & 2015.06\\
P$_{\textrm{b}}$ (yr) & 22.7 & 21.8 & 23.7 & 20.9 & 25.2 & 22.5 & 21.6 & 23.6 & 20.7 & 25.0\\
M$_{\textrm{b}}$ (M$_{\textrm{Jup}})$ & 8.35 & 5.76 & 10.91 & 3.19 & 13.42 & 8.03 & 5.41 & 10.64 & 2.80 & 13.20\\
a$_{\textrm{c}}$ (au) & 2.72 & 2.70 & 2.74 & 2.68 & 2.76 & 2.72 & 2.70 & 2.74 & 2.68 & 2.76\\
e$_{\textrm{c}}$ & 0.248 & 0.156 & 0.359 & 0.063 & 0.489 & 0.241 & 0.151 & 0.348 & 0.062 & 0.476\\
i$_{\textrm{c}}$ (deg) & 97.860 & 84.044 & 110.352 & 69.863 & 121.177 & 88.852 & 88.139 & 89.568 & 87.424 & 90.280\\
$\omega_{\textrm{c}}$ (deg) & 94.59 & 77.54 & 118.47 & 57.28 & 170.61 & 95.61 & 77.88 & 120.71 & 56.41 & 172.77\\
$\Omega_{\textrm{c}}$ (deg) & 36.471 & 21.411 & 51.757 & 4.743 & 69.401 & 32.016 & 31.307 & 32.726 & 30.601 & 33.435\\
T$_\textrm{0c}$ & 2013.17 & 2013.00 & 2013.36 & 2012.73 & 2013.79 & 2013.18 & 2013.00 & 2013.38 & 2012.71 & 2013.81\\
P$_{\textrm{c}}$ (yr) & 3.39 & 3.36 & 3.41 & 3.34 & 3.44 & 3.39 & 3.36 & 3.41 & 3.34 & 3.44\\
M$_{\textrm{c}}$ (M$_{\textrm{Jup}}$) & 9.37 & 8.44 & 10.43 & 7.56 & 11.79 & 9.18 & 8.31 & 10.14 & 7.47 & 11.31\\
M$_{*}$ & 1.76 & 1.73 & 1.79 & 1.70 & 1.82 & 1.76 & 1.74 & 1.79 & 1.71 & 1.82\\
$\alpha_{H0}^*$ (mas) & -0.095 & -0.253 & 0.065 & -0.410 & 0.228 & -0.025 & -0.151 & 0.101 & -0.275 & 0.228\\
$\delta_{H0}$ (mas) & 0.021 & -0.143 & 0.183 & -0.306 & 0.346 & -0.021 & -0.172 & 0.128 & -0.322 & 0.277\\
$\pi$ (mas) & 51.413 & 51.281 & 51.544 & 51.149 & 51.676 & 51.396 & 51.265 & 51.526 & 51.134 & 51.657\\
$\mu_{\alpha^*}$ & 4.965 & 4.947 & 4.981 & 4.928 & 4.997 & 4.963 & 4.948 & 4.977 & 4.932 & 4.992\\
$\mu_\delta$ & 83.967 & 83.948 & 83.986 & 83.925 & 84.004 & 83.969 & 83.950 & 83.987 & 83.930 & 84.004\\
$\gamma$ (m/s) & -26.1 & -51.9 & -3.7 & -88.7 & 19.1 & -25.3 & -51.2 & -2.8 & -86.9 & 20.0\\
\enddata
\end{deluxetable*}

\begin{figure}[ht]
\includegraphics[width=\columnwidth,trim=2cm 3cm 2cm 7cm]{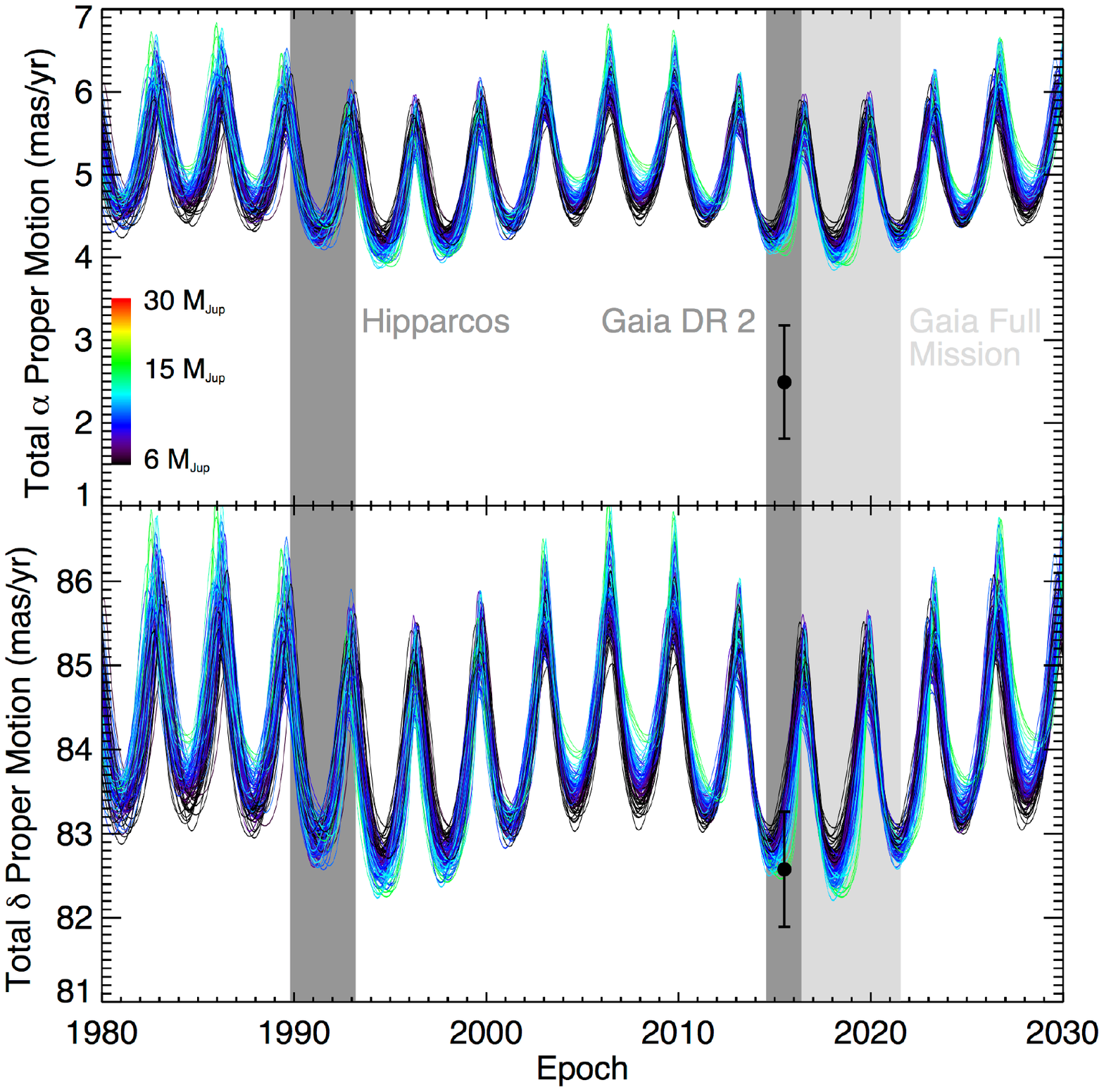}
\caption{Predicted proper motion from our two-planet coplanar fit.  As in the one-planet fit (Figure~\ref{fig:orbit_pm}), the $Gaia$ proper motion (which is not included in our fit) is a $\sim$2$\sigma$ outlier in right ascension.  The much shorter period of the oscillation, coupled with the larger amplitude, indicate that future $Gaia$ data releases could detect the astromeric signature of $\beta$ Pic c. \label{fig:coplanar_pm}}
\end{figure}

In the one-planet fit we found the $Gaia$ proper motion to be significantly offset ($\sim$2$\sigma$ from the predicted astrometric motion of the star in right ascension (Figure~\ref{fig:orbit_pm}).  Considering the proper motion of the star in the coplanar two-planet fit does not resolve this, as Figure~\ref{fig:coplanar_pm} shows this offset remains the same.  The two planets are similar in mass, but the inner planet accounts for more of the proper motion signature, since stellar orbital velocity scales as $a^{-0.5}$, and $\beta$ Pic c is $\sim$3.6 times closer to the parent star than $\beta$ Pic b.  The shorter period suggests that future $Gaia$ data releases could detect the astrometric motion of $\beta$ Pic due to the inner planet, as this orbit fit predicts significant acceleration of the star.

\subsection{Comparison to evolutionary models}

\begin{figure}[ht]
\includegraphics[width=\columnwidth,trim=2cm 3cm 2cm 7cm]{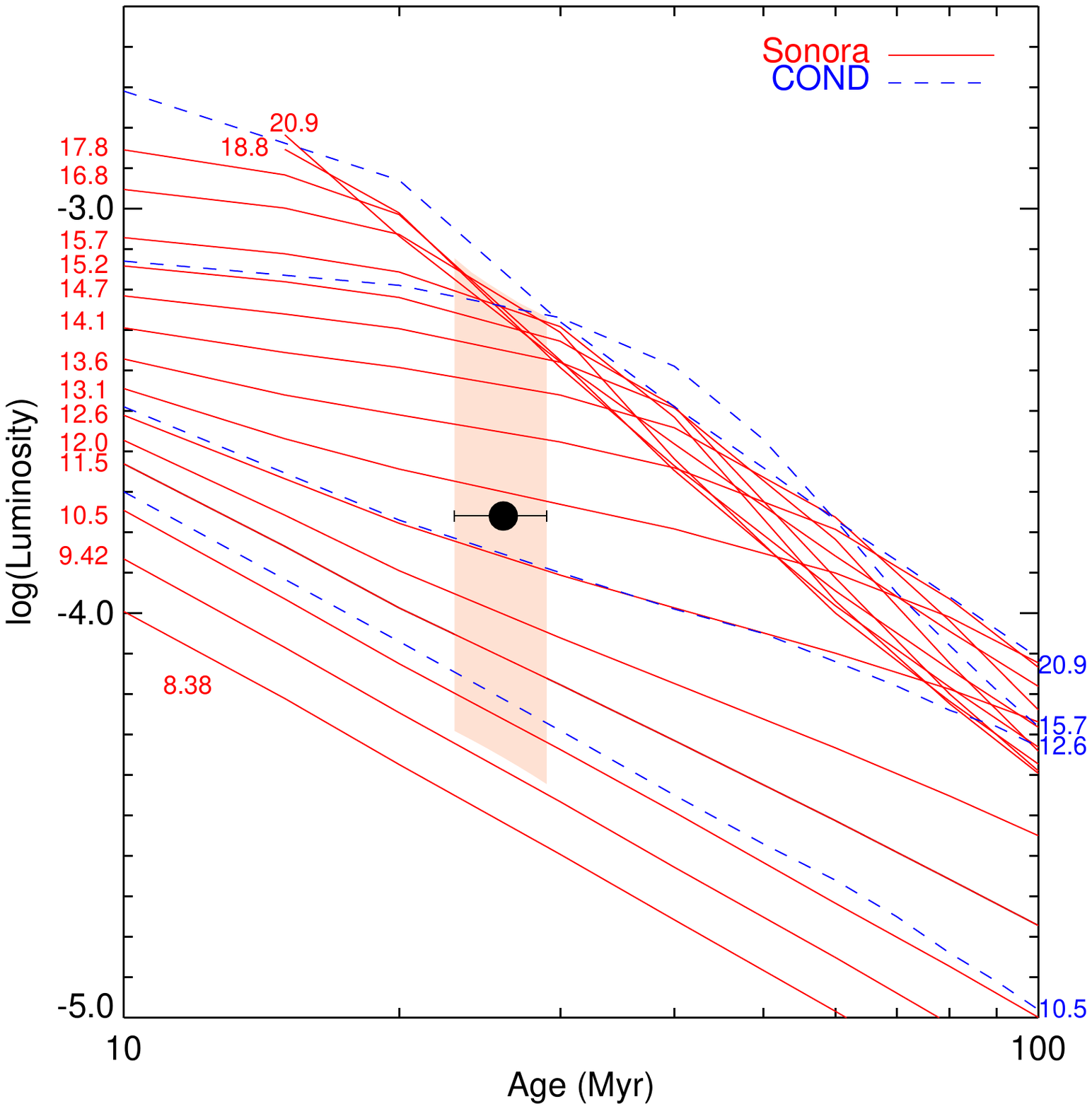}
\caption{Comparison of the luminosity of $\beta$ Pic b \citep{Chilcote:2017fv} and the age \citep{Nielsen:2016ct} to the Sonora (Marley et al. 2019 in prep) and COND \citep{baraffe03} models.  Luminosity is given in solar units, and red and blue numbers mark the masses of the tracks in M$_{\rm Jup}$ for the Sonora and COND models, respectively.  The light red shaded region represents the 1$\sigma$ region for the mass of the planet from our one-planet Case 3 fit and the Sonora models, consistent with the expected mass given the luminosity measurement. \label{fig:modelfig}}
\end{figure}

\begin{figure}[ht]
\includegraphics[width=\columnwidth,trim=2cm 3cm 2cm 7cm]{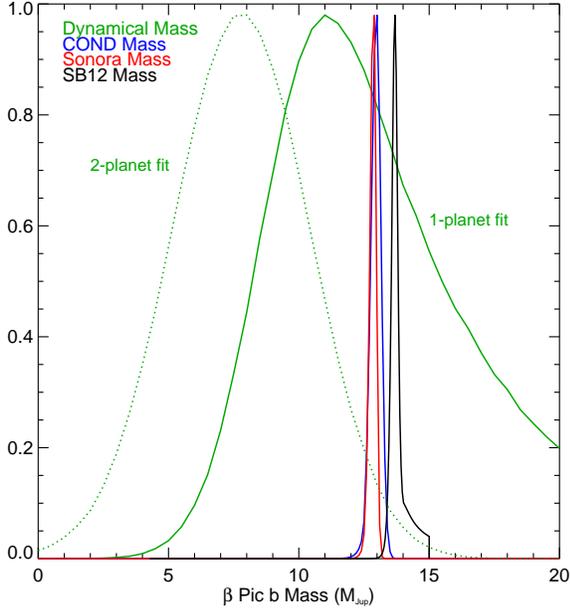}
\caption{Mass posteriors for $\beta$ Pic b from our orbit fit for the one-planet Case 3 fit (green solid curve) and the two-planet coplanar fit (green dotted line), compared to model-derived masses  from COND \citep{baraffe03}, Sonora (Marley et al. 2019 in prep), and SB12 \citep{Spiegel:2012} (hybrid clouds, solar metallicity) based on the luminosity and age of the planet.  Our one-planet fit dynamical mass is consistent with the model predictions, though with the current precision we cannot differentiate between the different models.  The two-planet fit mass is more discrepant with these model predictions, with less than a 5\% probability that the mass is larger 12.5 M$_\textrm{Jup}$. \label{fig:modelfig2}}
\end{figure}

\citet{Chilcote:2017fv} analyzed the SED of $\beta$ Pic b, and found a model-dependent mass of 12.9 $\pm$ 0.2 M$_{\rm Jup}$ using the bolometric luminosity of the planet, though this error bar does not include model uncertainty.  Figure~\ref{fig:modelfig} compares the luminosity determined by \citet{Chilcote:2017fv} of $\log{\frac{L}{L_\odot}} = -3.76\pm$ 0.02 to predictions from the COND \citep{Baraffe:2003bj} and Sonora (Marley et al. 2019 in prep) model grids, as well as the predicted luminosity given our dynamical mass measurement.  As expected, the \citet{Chilcote:2017fv} luminosity is significantly more precise than the uncertainty on our model-based prediction, given the $\sim$30\% errors on the dynamical mass, nevertheless the estimate is consistent with the measurement.  We compare our dynamical mass measurement to model predictions from this luminosity for the COND, Sonora, and SB12 \citep{Spiegel:2012} model grids in Figure~\ref{fig:modelfig2}, showing that the model-dependent luminosity estimates are well within the range implied from our one-planet fit mass measurement of 12.8$^{+5.5}_{-3.2}$ M$_{\rm Jup}$. While the three model-dependent mass estimates all have exquisite precision relative to the dynamical mass, they are in significant disagreement with one another. The hot-start models (COND and Sonora) predict a significantly lower mass for the planet than the highest entropy models from the \citet{Spiegel:2012} warm-start grid, and so the hot-start mass PDFs reach a maximum closer to the peak of the dynamical mass PDF than the warm-start PDF, though all three model PDFs have peaks within the 1$\sigma$ range of our dynamical mass measurement.  The two-planet fit mass for $\beta$ Pic b is significantly lower, with less than 5\% of orbits corresponding to a mass larger than 12.5 M$_\textrm{Jup}$, more in tensions with the model masses.

\textit{Gaia} DR 3 proper motions and accelerations, along with continued monitoring of the relative orbit by direct imaging, will likely further constrain the orbit and the mass, and the DR 4 intermediate data will allow for a full fit including individual absolute astrometric measurements from \textit{Hipparcos} and \textit{Gaia} and ground-based relative astrometry and radial velocities.
A precise determination of the mass of the planet using these data will allow $\beta$ Pic b to be used as an empirical calibrator for evolutionary models at young ages where planets are still significantly radiating away their formative heat. We note that the luminosity-derived masses discussed previously assume prompt planet formation. A delay between star and planet formation may lead to a significantly younger age for $\beta$ Pic b than its host star (e.g., \citealp{Currie:2009ee}). Our current constraints on the dynamical mass of the planet do not allow us to distinguish between a prompt and delayed formation scenario assuming a given evolutionary model.  The 8.0$\pm 2.6$ M$_\textrm{Jup}$ mass from the two-planet fit would require a significantly delayed epoch of planet formation to bring the luminosity in line with evolutionary models. A precise, model-dependent measurement of the entropy of formation will greatly constrain formation models for wide-separation giant planets as well.

\begin{figure}[ht]
\includegraphics[width=\columnwidth,trim=2cm 3cm 2cm 7cm]{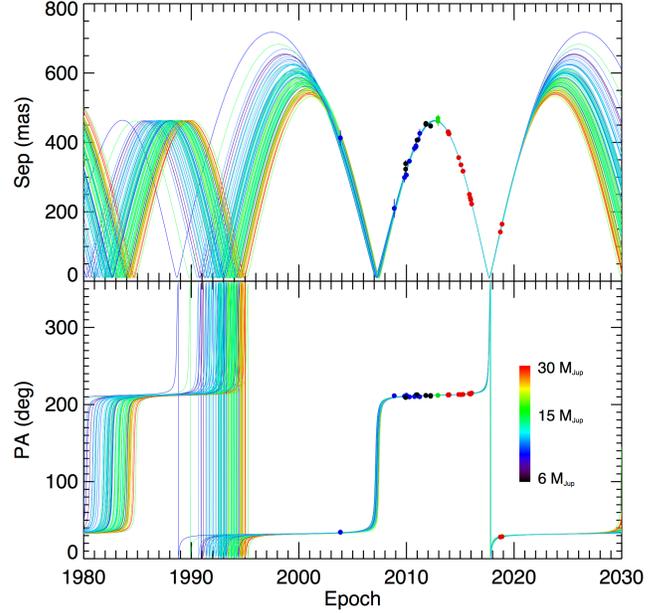}
\caption{Future tracks of the Case 3, one-planet fit $\beta$ Pic b orbit, again color-coded by mass, drawn from the posterior of the orbit including all imaging data except SPHERE, and \textit{Hipparcos} and \textit{Gaia}.  There is a general trend where higher mass planets result in a faster turn-around in $\sim$2024.\label{fig:track2}}
\end{figure}

\begin{figure}[ht]
\includegraphics[width=\columnwidth,trim=2cm 3cm 2cm 7cm]{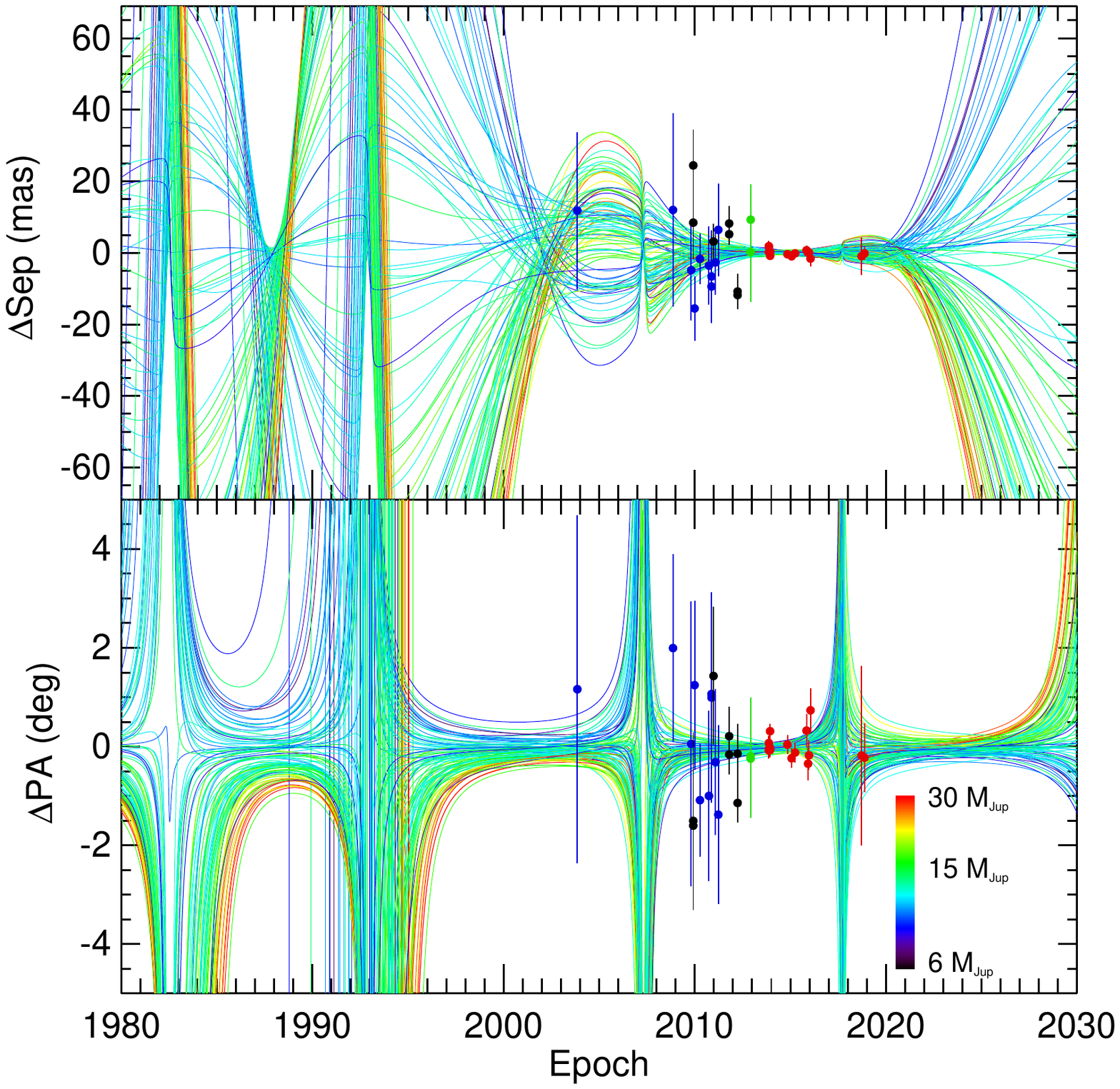}
\caption{Same as Figure~\ref{fig:track2}, but with the lowest $\chi^2$ orbit subtracted from the tracks to give more detail.   Further monitoring of the system between 2020-2022 at the 1 mas level will greatly reduce the uncertainty in the orbital parameters, particularly in period.\label{fig:track3}}
\end{figure}

In the meantime, more ground-based relative astrometry will also increase the mass precision.  Figure~\ref{fig:triangle} shows significant covariance between eccentricity, period, and planet mass.  Thus, further constraints on the orbital parameters will reduce the mass errors.  Figure~\ref{fig:track2} shows a significant divergence in the orbit tracks beyond $\sim$2022, with higher masses generally corresponding to the shortest orbital periods.

To highlight this dependence, in Figure~\ref{fig:track3} we subtract off the lowest $\chi^2$ orbit from each of the tracks.  The prediction for separation at 2020.0 has an uncertainty of 1.8 mas, which rises to 3.5 mas at 2021.0, and 8.2 mas at 2022.0.  In comparison, \citet{Wang:2016gl} demonstrated the ability to reach relative astrometric precision of less than 1 mas on $\beta$ Pic b with GPI when the separation was above $\sim$230 mas, a separation the planet should have reached again in June 2019.  Thus, continued monitoring with GPI and SPHERE will further refine the orbit determination.  While there is not a direct correlation between mass and separation between 2020-2022, if the shortest orbital periods are ruled out, this will also rule out the largest values of planet mass, leading to a more precise mass measurement.

\begin{figure}[ht]
\includegraphics[width=\columnwidth,trim=2cm 3cm 2cm 7cm]{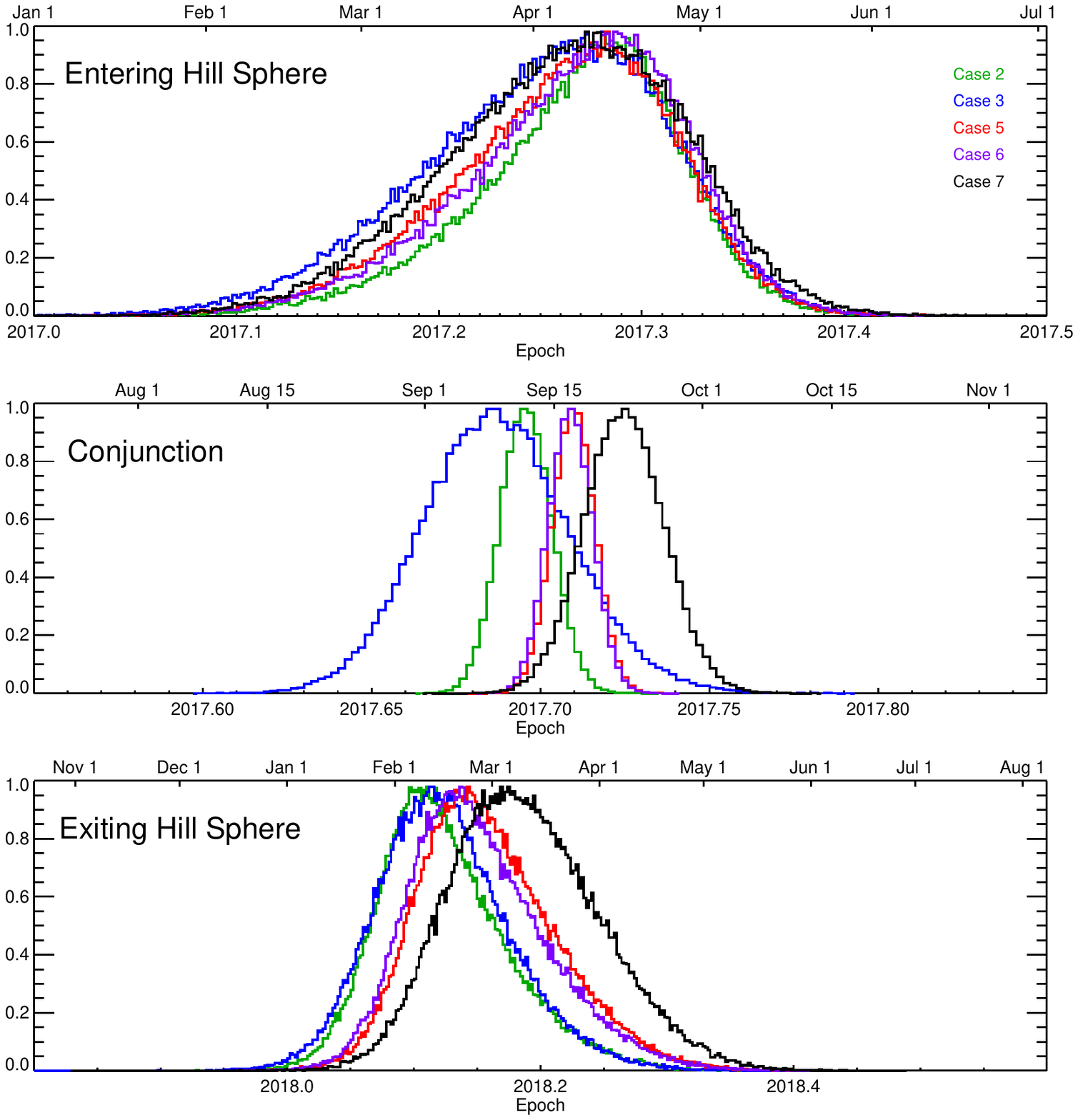}
\caption{Posteriors on the Hill sphere crossing and conjunction (closest approach) between 2017 and mid-2018, for orbit fits including data from 2018.  For most models the Hill sphere crosses in front of the star in mid-April 2017, and the crossing lasts until early-February 2018.  In all fits including the 2018 GPI data, conjunction occurs in a ten-day window between 2017-9-11 and 2017-9-20 (2$\sigma$).
\label{fig:hillsphere}}
\end{figure}

\subsection{Disk and Hill sphere}

Our new constraints for the orbital geometry of $\beta$ Pic b are also relevant to the ongoing investigations of planet--disk dynamical interactions. Periastron occurs near maximum elongation close to the sky plane in the SW at epoch 2013.72.  With $a$ = 10.18 au and $e$ = 0.122, the projected periastron and apastron separations are 8.94 au and 11.42 au, respectively. If the planet is secularly forcing the eccentricities of the nearby material \citep{wyatt:1999}, then the inner cavity cleared by the planet should have a stellocentric offset similar to that of Fomalhaut's dust belt \citep{kalas:2005} of roughly 4.18 au or 215 mas.  The current scattered light data \citep{golimowski:2006,apai:2015} and millimeter continuum maps \citep{matra:2019} do not have the required angular resolution to directly detect the hypothesized offset.  However, the $\sim20\%$ stronger mid-infrared thermal emission from the SW side of the disk compared to the NE is consistent with the offset \citep{lagage:1994, wahhaj:2003}.

Given the edge-on nature of the $\beta$ Pic debris disk and the planet's orbit, as well as evidence for a transit-like event in 1981 \citep{LecavelierDesEtangs:2009jt}, determining if the planet transits became of great interest (e.g. \citealt{Nielsen:2014}, \citealt{Wang:2016gl}).  \citet{Wang:2016gl} ruled out the prospect of the planet itself transiting at 10 $\sigma$, and from our Case 3, we find the closest approach by the planet to be a projected separation of 22.7 $\pm$ 1.9 R$_*$, where we take the radius of the star to be 1.8 R$_\odot$, from the interferometric measurement of \citet{difolco:2004} of 1.8$\pm$ 0.2 R$_\odot$ (Figure~\ref{fig:hillsphere}).  From our MCMC chain, the smallest projected separation reached is 13.6 R$_*$ (0.11 au, 0.09 $r_H$), similar to all of our other cases (minimum values of 13.5--14.3 R$_*$) except Case 7, which utilized only imaging datasets from NACO and SPHERE.  In this orbit fit, we find a minimum projected separation of 17.3 $\pm$ 3.1 R$_*$, with a single value out of 10$^6$ having a separation $<$5 R$_*$, at 4.6 R$_*$.  Thus, we concur with \citet{Wang:2016gl} that the astrometry strongly disfavors transit, at the 12.2 $\sigma$ level for Case 3.  This conclusion is the same for the two-planet fit and the three-planet fit, with the smallest projected separation barely changing, from $22.7 \pm 1.9$ R$_*$ for Case 3 to $23.0 \pm 1.9$ R$_*$ for both the unconstrained mutual inclination and coplanar fits.

While transit of the planet itself is ruled out, the Hill sphere of the planet passes in front of the star, as noted by \citet{Wang:2016gl}. From 2017 to 2018, this offered a rare opportunity to probe the circumplanetary environment of a young Jovian exoplanet at large orbital separations where the influence of the star is minimal. There have been numerous observational efforts to monitor $\beta$ Pic both photometrically and spectroscopically during this Hill sphere crossing (e.g. \citealt{Mekarnia2017}, \citealt{Stuik2017}, \citealt{kalas:2019}, \citealt{mellon:2019}), so pinpointing the timeframe of the crossing is of prime interest to put these monitoring programs into context.  

The Hill sphere radius ($r_H$) is given by

\begin{equation}
    r_H \approx a (1-e) \left (\frac{M_P}{3M_*} \right)^\frac{1}{3}
\end{equation}

\noindent a function of the semi-major axis ($a$), eccentricity ($e$), planet mass ($M_P$), and star mass ($M_*$), from \citet{hamilton:1992}.  For our Case 3 orbit fits, we find a value of the Hill sphere radius of 1.18$^{+0.15}_{-0.11}$ au.  We find that the Hill sphere first crosses in front of the star 2017-4-11 ($\pm$ 18 days), and the crossing lasts until 2018-2-16 ($\pm$ 18 days).  Conjunction is more tightly constrained, taking place on 2017-9-13 ($\pm$2.8 days).  The uncertainty in the timing of the Hill sphere crossings is dominated by the error on the planet mass; fixing the planet mass results in timing windows $\sim$6x smaller, more in line with the precision on time of conjunction.  The predictions for Hill Sphere crossing does change in the two-planet fit, largely because the size of the Hill sphere is reduced thanks to a smaller planet mass for $\beta$ Pic b, starting in 2017-5-5 ($\pm$ 18 days) and ending 2018-1-24 ($\pm$ 18 days).  Conjunction changes slightly for the two-planet fit, 2017-9-15 ($\pm$2.9 days) for the unconstrained mutual inclination fit, and 2017-9-14 ($\pm$2.8 days) for the coplanar fit.

The dates of these events vary from the different orbit fits, as shown in Figure~\ref{fig:hillsphere}.  For Case 2, where the planet mass is not determined from the fit, we randomly sample from the Case 3 planet mass posterior.  Generally, posteriors are similar for datasets that include the GPI 2018 astrometry (Cases 2, 3, 5, and 6).  For example, time of conjunction has a median date of 2017-9-12 and 2017-9-13 for Cases 2 and 3, 2017-9-17 for Cases 5 and 6, and 2017-9-28 for Case 7.  The large offset for the Case 7 fit, using only NACO and SPHERE data, suggests there is a bias between GPI and SPHERE relative astrometry, either due to instrumental calibration or data pipeline systematics.  Indeed, for time of conjunction, the fits combining GPI and SPHERE data are in between fits excluding one of the two instruments.  Our Case 6 combines data from the two instruments with offset terms in separation and position angle, however the posteriors on time of conjunction and Hill sphere crossings are identical whether these offests are applied (case 6) or not (Case 5).  Thus a single offset between the two instruments does not appear to address the issue, suggesting either a different parameterization is needed, or the bias is time-variable.  Further work is needed to understand how this offset arises between the two instruments.  Nguyen et al. 2019 \textit{submitted} is currently analyzing multiple epochs of the same calibration field taken over GPI's lifetime as a validation of the astrometric calibration presented in \citet{derosa2019_north}.

\section{Conclusion}
We combine relative astrometry of the planet $\beta$~Pic~b with \textit{Gaia} postion and \textit{Hippacos} Intermediate Astrometric Data to refine the orbit and measure the mass of the planet.  We find a model-independent mass for the planet of 12.8$^{+5.5}_{-3.2}$ M$_{\rm Jup}$, consistent with predictions from hot-start evolutionary models given the luminosity of the planet and age of the system.  We find significant evidence for non-zero, but low, eccentricity for the planet, finding a value of 0.12$^{+0.04}_{-0.03}$.  Our comparison to previous work by \citet{snellen:2018} and \citet{dupuy:2019} underscores the importance of performing a joint fit to the space-based absolute astrometry and ground-based relative astrometry.  The reason for the offset between our fit and that of \citet{dupuy:2019} is less clear, and could be a combination of new relative astrometry, their fitting additional radial velocity data of the star and planet, and their use of recalculated \textit{Hipparcos} and \textit{Gaia} catalog values.

When including the radial velocities of the star from \citet{lagrange:2019} and adding an additional planet to the fit, $\beta$ Pic c, we find a significantly lower mass for $\beta$ Pic b, $8 \pm 2.6$ M$_\textrm{Jup}$, and no significant difference in the orbital parameters whether the planets are assumed to be coplanar or not.  We predict significant astrometric motion of the star from the orbit of $\beta$ Pic c, and future $Gaia$ data releases may be able to detect the signature of this inner planet.

We have constrained the time of conjunction of the planet to an accuracy of 2.7 days, and the Hill sphere entrance and exit to 18 days.  These values will guide analysis of the photometric monitoring of the star over the last two years to search for circumplanetary material transiting in front of the star.

Future monitoring of $\beta$ Pic by both ground-based imaging and \textit{Gaia} should further improve the precision on the measurement of the planet mass.  As the planet moves further from the star, GPI and SPHERE will be able to determine the relative astrometry with increasing precision.  Similarly, if \textit{Gaia} astrometry on bright stars can be improved, the reflex motion of the star over the \textit{Gaia} mission can be used to directly constrain the planet mass.

The combination of directly-imaged short-period substellar companions and precision \textit{Gaia} astrometry represents a new opportunity to directly measure the masses of these objects.  As shown in the case of $\beta$ Pic, strong constraints on the orbital parameters allows us to connect the motion from the $\sim$1991 \textit{Hipparcos} IAD directly to the $\sim$2015 \textit{Gaia} astrometry.  Further \textit{Gaia} data releases and ground-based imaging will allow us to measure, or set upper limits on, other directly imaged substellar companions with shorter orbital periods, including 51 Eridani b \citep{Macintosh:2015ew}, HR 2562 \citep{Konopacky:2016dk}, and HD 984 \citep{Meshkat:2015hd}.  When coupled with JWST mid-IR observations of these objects that will sample the part of the SED with the bulk of the flux, we can directly compare the luminosity predictions of evolutionary models to the measured masses from absolute and relative astrometry.

\acknowledgements

Based on observations obtained at the Gemini Observatory, which is operated by the Association of Universities for Research in Astronomy, Inc., under a cooperative agreement with the NSF on behalf of the Gemini partnership: the National Science Foundation (United States), the National Research Council (Canada), CONICYT (Chile), Ministerio de Ciencia, Tecnolog\'{i}a e Innovaci\'{o}n Productiva (Argentina), and Minist\'{e}rio da Ci\^{e}ncia, Tecnologia e Inova\c{c}\~{a}o (Brazil). This research has made use of the SIMBAD and VizieR databases, operated at CDS, Strasbourg, France. This work has made use of data from the European Space Agency (ESA) mission {\it Gaia} (\url{https://www.cosmos.esa.int/gaia}), processed by the {\it Gaia} Data Processing and Analysis Consortium (DPAC, \url{https://www.cosmos.esa.int/web/gaia/dpac/consortium}). Funding for the DPAC has been provided by national institutions, in particular the institutions participating in the {\it Gaia} Multilateral Agreement.
 This research used resources of the National Energy Research Scientific Computing Center, a DOE Office of Science User Facility supported by the Office of Science of the U.S. Department of Energy under Contract No. DE-AC02-05CH11231.  This work used the Extreme Science and Engineering Discovery Environment (XSEDE), which is supported by National Science Foundation grant number ACI-1548562.
 
 J.R. and R.D. acknowledge support from the Fonds de Recherche du Qu\'{e}bec. JRM's work was performed in part under contract with the California Institute of Technology (Caltech)/Jet Propulsion Laboratory (JPL) funded by NASA through the Sagan Fellowship Program executed by the NASA Exoplanet Science Institute. Support for MMB's work was provided by NASA through Hubble Fellowship grant \#51378.01-A awarded by the Space Telescope Science Institute, which is operated by the Association of Universities for Research in Astronomy, Inc., for NASA, under contract NAS5-26555. Supported by NSF grants AST-1411868 (E.L.N., K.B.F., B.M., J.P., and J.H.), AST-141378 (G.D.), and AST-1518332 (R.D.R., J.J.W., T.M.E., J.R.G., P.G.K.). Supported by NASA grants NNX14AJ80G (E.L.N., B.M., F.M., and M.P.), NNX15AC89G and NNX15AD95G (B.M., T.M.E., R.J.D.R., G.D., J.R.G., P.G.K.), NN15AB52l (D.S.), NSSC17K0535 (E.L.N., R.D.R., B.M., J.-B.R.). J.J.W. is supported by the Heising-Simons Foundation 51~Pegasi~b postdoctoral fellowship. Portions of this work were performed under the auspices of the U.S. Department of Energy by Lawrence Livermore National Laboratory under Contract DE-AC52-07NA27344. This work benefited from NASA's Nexus for Exoplanet System Science (NExSS) research coordination network sponsored by NASA's Science Mission Directorate.

\facility{Gemini:South (GPI)}
\software{Astropy \citep{astropy}, emcee \citep{ForemanMackey:2013io}, GPI DRP \citep{Perrin:2016gm}, IDL Astronomy Library \citep{idlastro}, pyKLIP \citep{Wang:2015th}}

\newpage

\bibliographystyle{aasjournal}   
\bibliography{main} 

\end{document}